\documentclass[journal]{IEEEtran}
\usepackage{amssymb}
\usepackage{color,graphicx}
\usepackage[noadjust]{cite}
\usepackage[cmex10]{amsmath}
\usepackage{amsopn}
\usepackage{stfloats}
\usepackage{multirow}
\usepackage{bigstrut}
\usepackage{enumerate}
\usepackage{float}
\usepackage{accents}

\usepackage{pifont}

\usepackage[table]{xcolor}

\newtheorem{definition}{Definition}
\newtheorem{proposition}{Proposition}

\newtheorem{corollary}{Corollary}
\newtheorem{remark}{Remark}
\newtheorem{example}{Example}
\newcommand{\remarkend}{\IEEEQEDopen}

\allowdisplaybreaks

\begin{document}
\title{Cooperative Multi-Sender Index Coding}
\author{Min Li,~\IEEEmembership{Member,~IEEE}, Lawrence Ong,~\IEEEmembership{Member,~IEEE}, and Sarah J. Johnson,~\IEEEmembership{Member,~IEEE}
\thanks{This work was supported by the Australian Research Council under Grant FT140100219 and Grant DP150100903. This paper was presented in part at the IEEE International Symposium on Information Theory, Aachen, Germany, June 25-30, 2017.
\par Min Li was with the University of Newcastle, and is now with the School of Engineering, Macquarie University, Sydney, NSW~2109, Australia (email: min.li@mq.edu.au). Lawrence Ong and Sarah J. Johnson are with the School of Electrical Engineering and Computing, The University of Newcastle, Callaghan, NSW~2308, Australia (e-mail: lawrence.ong@newcastle.edu.au, sarah.johnson@newcastle.edu.au).
\par Copyright (c) 2017 IEEE. Personal use of this material is permitted. However, permission to use this material for any other purposes must be obtained from the IEEE by sending a request to pubs-permissions@ieee.org.}}

\maketitle

\begin{abstract}
In this paper, we propose a new coding scheme and establish new bounds on the capacity region for the multi-sender unicast index-coding problem. We revisit existing partitioned Distributed Composite Coding (DCC) proposed by~Sadeghi~\textit{et al.} and identify its limitations in the implementation of multi-sender composite coding and in the strategy of sender partitioning. We then propose two new coding components to overcome these limitations and develop
a multi-sender Cooperative Composite Coding (CCC). We show that CCC can strictly improve upon partitioned DCC, and is the key to achieve optimality for a number of index-coding instances. The usefulness of CCC and its special cases is illuminated via non-trivial examples, and the capacity region is established for each example. Comparisons between CCC and other non-cooperative schemes in recent works are also provided to further demonstrate the advantage of CCC.
\end{abstract}

\begin{IEEEkeywords}
Composite coding, cooperative compression, index coding, random binning.
\end{IEEEkeywords}

\section{Introduction}
\par Index coding addresses efficient broadcast problems where multiple receivers each wish to decode some messages from a common channel, and they each know a subset of messages \textit{a priori}. Originally motivated from satellite communications~\cite{birk1998informed,bar2011index}, the index-coding problem is shown to have rich connections with network coding~\cite{el2010index,effros2015equivalence}, coded caching~\cite{maddah2014fundamental}, distributed storage~\cite{mazumdar2014duality,arbabjolfaei2015three} and topological interference management in wireless communications~\cite{jafar2014topological}.

\subsection{Background and Related Works}

\par In the classic index-coding setup, one sender encodes a set of messages and broadcasts the codeword to multiple receivers through a noiseless channel. Each message is requested by only one receiver, and each receiver requests only one message. The aim is to find the optimal broadcast rate (i.e., the normalized codeword length) or the capacity region such that each receiver can correctly decode what it wants given what it knows. This problem is referred to as the single-sender unicast index-coding problem~\cite{birk1998informed,lubetzky2009nonlinear,bar2011index,chaudhry2011complementary,shanmugam2013local,arbabjolfaei2013capacity,arbabjolfaei2014index,yu2014duality,arbabjolfaei2015structural,thapa2016interlinked,ong2016optimal}.

\par Most studies above cast the problem into side-information digraphs and derived bounds on the broadcast rate using graph-related quantities, such as the size of maximum acyclic induced subgraph (MAIS) for the lower bound~\cite{bar2011index}, and the min-rank over digraph-induced matrices~\cite{bar2011index}, the (partial) clique-covering number~\cite{birk1998informed}, the cycle-covering number~\cite{chaudhry2011complementary}, and the local chromatic number~\cite{shanmugam2013local} for upper bounds. Recently, Thapa~\textit{et~al.}~\cite{thapa2016interlinked} introduced the interlinked-cycle structure (a general form of overlapping cycles) and proposed an interlinked-cycle-cover (ICC) scheme using scalar linear index codes. It has been proved that the ICC scheme can outperform all schemes above for certain digraphs. {The optimality of each scheme mentioned was proved for certain classes of graphs in each work, but none of them is optimal in general, since all these schemes give linear index codes and there exist some graphs for which non-linear index coding can outperform the linear optimum~\cite{lubetzky2009nonlinear}}.

\par Arbabjolfaei~\textit{et~al.}~\cite{arbabjolfaei2013capacity,arbabjolfaei2014index} instead viewed the problem information-theoretically and developed bounds using techniques in network information theory such as random coding (binning). {In particular, they devised a layered random coding scheme, termed \textit{composite coding}, for the achievability. In the scheme, each non-empty subtuple of the messages at the sender is referred to as a \textit{composite message}}. The sender first encodes each composite message into a composite index at an appropriate rate by random binning, and then uses flat coding to encode the composite indices and broadcast to the receivers, while upon reception, each receiver leverages its side information, to first retrieve all composite indices, and then to decode the desired messages from the relevant composite indices. Composite coding is shown to achieve the optimal broadcast rate~\cite{arbabjolfaei2014index} and the information capacity region~\cite{arbabjolfaei2013capacity} asymptotically (as the message size tends to infinity) for all unicast index-coding problems with five or fewer receivers. However, this composite coding is still not optimal in general, as shown by a 6-receiver instance~\cite{thapa2016interlinked} and a 7-receiver instance~\cite{jafar2014topological}.

\par Single-sender multicast index-coding problems (where each message may be requested by more than one receiver) have also been studied in the literature using both graph theory~\cite{chaudhry2011complementary,ong2012optimal,blasiak2013broadcasting,neely2013dynamic} and rate distortion theory~\cite{Unal2016}. In particular, the optimal broadcast rate was established~\cite{Unal2016} for a number of multicast scenarios, including all instances with up to three receivers. But the general multicast/unicast index-coding problem has been shown to be NP-hard~\cite{chaudhry2011complementary}.

\par In the setups above, all messages are assumed to be stored and transmitted by a central sender. However, in various scenarios of interest, messages might be distributed across multiple senders, and users might be served jointly by these senders. For instance, in storage networks, data can be distributed over multiple storage locations. In satellite communications, multiple satellites with local messages might jointly serve multiple clients on the downlink for better coverage. As a third example, in the video-content-driven 5G heterogenous networks~\cite{shanmugam2013femtocaching}, caching parts of video content (i.e., messages) at distributed small-cell base stations and at user devices during off-peak hours has great potential in alleviating the load of the backhaul network and reducing end-to-end transmission delay during peak hours. Assume that video content is already properly cached. Unleashing the potential of cached video calls for design of efficient transmission schemes from multiple senders with cached-messages to multiple users each with some side-information, i.e., a general multi-sender index-coding problem.

\par Ong~\textit{et~al.}~\cite{ong2016multi}~were the first to investigate an instance of such a setting. In particular, they proposed the joint use of \textit{information-flow graph} and \textit{message graph} to represent a multi-sender index-coding problem, and developed lower and upper bounds on the optimal broadcast rate for the \textit{multicast single-uniprior} instances. Within this class of instances, it was further shown that the bounds coincide for the special case where no two distinct senders have any messages in common. More recently, Thapa~\textit{et~al.}~\cite{thapa2016twosender} extended the single-sender version of the cycle-cover, clique-cover and local-chromatic number schemes to the two-sender unicast problem, and established the optimal broadcast rate under certain combinations of side-information graph and message graph.

\par Sadeghi~\textit{et~al.}~\cite{sadeghi2016distributed} considered a general multi-sender unicast setting, where there are $2^N-1$ senders each containing a different subset of the $N$~messages in the system and each connected to all $N$~receivers by noiseless broadcast links of arbitrary finite capacity. Both inner and outer bounds on the capacity region have been proposed for the problem studied. {In particular, the inner bound was attained by a distributed version of aforementioned composite coding~\cite{arbabjolfaei2013capacity}, termed \textit{partitioned Distributed Composite Coding} (DCC). In the scheme, senders are partitioned into \textit{disjoint} sender-groups, any message that appears in different sender-groups is split at appropriate rates, and a composite coding problem is then formed and solved for each sender-group.} It was indicated that partitioned DCC with all senders in the same group suffices to achieve the capacity region for all non-isomorphic index-coding instances with $N=3$ messages for arbitrary link capacities~\cite[Section IV.B]{sadeghi2016distributed}. In addition, partitioned DCC with appropriate sender grouping was shown to be useful to achieve the sum-rate outer bound through an example with $N=4$ messages and $15$ senders each with unit link capacity\footnote{We would like to point out that, in fact, partitioned DCC can achieve the entire capacity region for this particular example, if the second and the fifth groups~\cite[Table II]{sadeghi2016distributed} are combined as a single group (see discussions in \textit{Example~\ref{example:parastoo}} here). The converse proof needs customized Shannon-type inequalities as shown in Appendix~\ref{appendix:converse:parastoo} in this paper.}~\cite[Section IV.C]{sadeghi2016distributed}. However, we show that for $N=4$ messages, partitioned DCC can be sub-optimal even for an index-coding instance with only two senders, meaning that the link capacities associated with the remaining senders in the system are set to zero ({see \textit{Example~\ref{example:DCC:fails}} later in Section~\ref{sec:cooperative:compression}}).

{\begin{table*}[t]
\centering
  \caption{A Summary of Coding Schemes Proposed in This Work and Existing Partitioned DCC}
    \begin{tabular}{|c|c|c|c|c|}
    \hline
    {Rate Region~/~Techniques} & {Composite-Message Compression} &{Sender Grouping}&{Proposed by} \bigstrut\\
    \hline
    {${\mathcal R}~({\mathcal R}_{\text{DCC-a}})$} & {non-cooperative} & {all senders in a group} &{Sec. IV.B,~\cite{sadeghi2016distributed}} \bigstrut\\
    \hline
    {${\mathcal R}_{\text{S}}~(\text{a.k.a.}~{\mathcal R}_{\text{DCC}})$} & {non-cooperative} & {sender partitioning} &{Sec. IV.C,~\cite{sadeghi2016distributed}} \bigstrut\\
    \hline
 \cellcolor{blue!10} {${\mathcal R}_{\text{LS}}$} & {non-cooperative} & \textcolor{red}{joint link-and-sender partitioning} & {this work}; ~\cite{liu2017distributed} \bigstrut\\
\hline
\cellcolor{purple!10} {${\mathcal R}_{\text{C}}$} & {\textcolor{red}{cooperative}} & {all senders in a group} & \multirow{3}[5]{*}{{this work}} \bigstrut \bigstrut\\
\cline{1-3}    \cellcolor{purple!10} {${\mathcal R}_{\text{CS}}$} & {\textcolor{red}{cooperative}} & {sender partitioning} & \bigstrut\\
\cline{1-3}    \cellcolor{purple!10} {${\mathcal R}_{\text{CLS}}$} & {\textcolor{red}{cooperative}} & \textcolor{red}{joint link-and-sender partitioning} &  \bigstrut\\
    \hline
    \end{tabular}%
\label{table:listofschemes}%
\end{table*}

\begin{figure*}[t]
\centering
\includegraphics[width=0.8\textwidth]{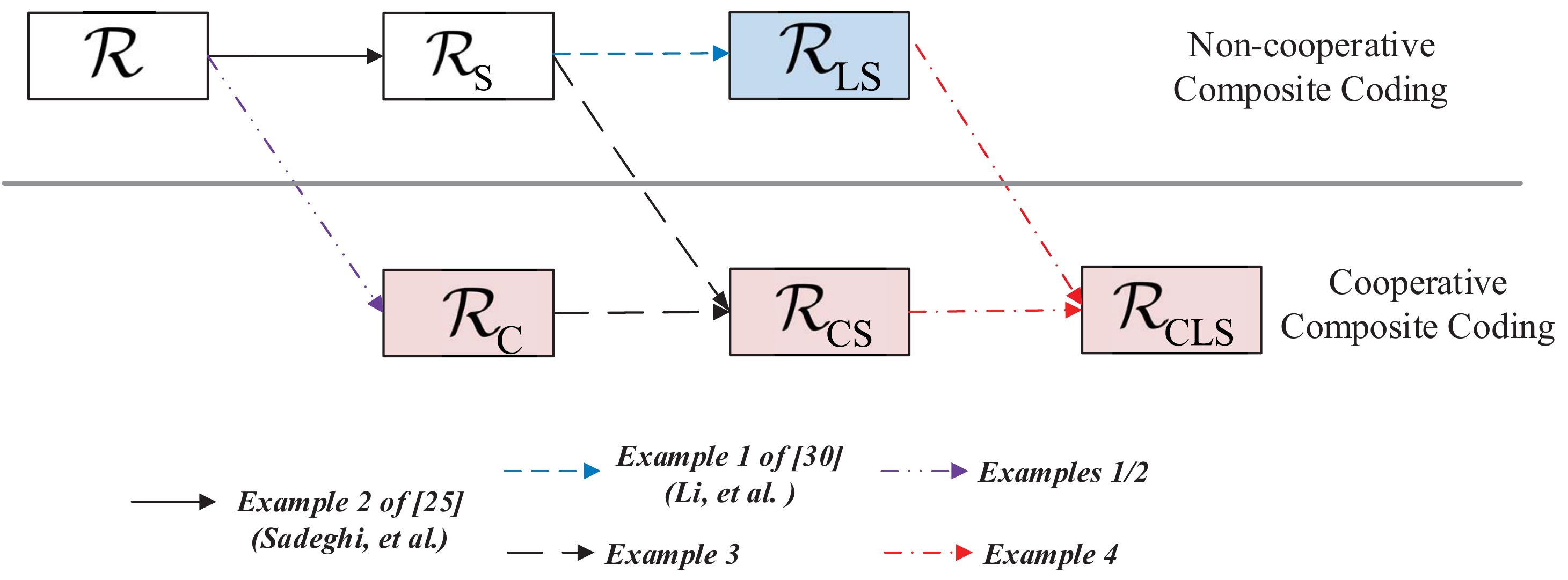}
\caption{Performance comparison of different schemes: The arrow from ${\mathcal R}_i$ to ${\mathcal R}_j$ indicates that ${\mathcal R}_i \subseteq {\mathcal R}_j$. {All the inclusions have been shown to be strict in the mentioned examples}. For notational convenience and if no ambiguity is caused, we simply use ``{${\mathcal R}_i \subsetneq {\mathcal R}_j$}" to denote the relationship. Note that this relationship is transitive, i.e., whenever {${\mathcal R}_i \subsetneq {\mathcal R}_j$} and {${\mathcal R}_j \subsetneq {\mathcal R}_k$, then also ${\mathcal R}_i \subsetneq {\mathcal R}_k$}.} \label{fig:scheme:comparison}
\end{figure*}}

\subsection{Our Contributions}

\par In this paper, we develop a new coding scheme and establish new bounds on the capacity region for the multi-sender index-coding problem.
\par Specifically, we revisit partitioned DCC and identify its limitations in the implementation of multi-sender composite coding and in the strategy of sender partitioning. Two techniques are proposed to overcome these limitations. In the first technique, termed \textit{cooperative compression} of composite messages, all senders in each sender group encode the same composite message to the same composite index. We will show that this can only enlarge the composite rate region compared to its non-cooperative counterpart in partitioned DCC. The second technique, termed \textit{joint link-and-sender partitioning}~\footnote{Independent of and in parallel with this work, Liu~\textit{et~al.}~\cite{liu2017distributed} proposed an enhanced DCC scheme, which includes joint link-and-sender partitioning (the second technique proposed in this paper) as a special case. We will discuss this more in Section~\ref{sec:discussions}.}, generalizes sender partitioning aforementioned in partitioned DCC, where we allow each sender to split its link capacity and participate in different sender groups if necessary to enlarge the achievable rate region. Combining these two techniques, we develop a multi-sender \textit{Cooperative Composite Coding} (CCC) and prove that it can strictly outperform partitioned DCC.

\par {As listed in Table~\ref{table:listofschemes}, we also consider all possible schemes with different combinations of $(i)$ non-cooperative\footnote{{In contrast to CCC, partitioned DCC compresses the same composite message to different composite indices at different senders, and therefore we say that it is non-cooperative compression of composite messages.}} or cooperative compression of composite messages, $(ii)$ with all senders in a group, or with sender partitioning or with joint link-and-sender partitioning, aiming to illuminate the utility of each technique component. One category includes the general CCC (with rate region ${\mathcal R}_{\text{CLS}}$) and its two special cases with rate region ${\mathcal R}_{\text{C}}$, ${\mathcal R}_{\text{CS}}$, respectively, where the subscript is used to indicate the technique components each scheme has. The other category uses non-cooperative compression of composite coding. It includes partitioned DCC and its special case, whose rate region is denoted by ${\mathcal R}_{\text{S}}$ (a.k.a. ${\mathcal R}_{\text{DCC}}$), and ${\mathcal R}$ (or ${\mathcal R}_{\text{DCC-a}}$), respectively. It also includes a modified DCC scheme we propose (see Section~\ref{sec:discussions}), whose rate region is denoted by ${\mathcal R}_{\text{LS}}$. {The special cases of each scheme suffice to achieve the capacity region for certain instances (that is, certain link capacities and side information setups)}, as demonstrated in Section~\ref{sec:cooperative:compression}. The performance comparison of different schemes is depicted in Fig.~\ref{fig:scheme:comparison}. Note that the non-cooperative scheme is not a special case of the cooperative scheme, and so the arrows across category are not trivial to establish. Within the same category, while inclusions are trivial, as a scheme always include the schemes to its left as special cases, we will show strict inclusions with examples.}

\par The rest of the paper is organized as follows. Section~\ref{sec:model} formalizes the multi-sender index coding problem considered. Section~\ref{sec:preliminaries} revisits the key ideas and some related results in~\cite{sadeghi2016distributed}. Section~\ref{sec:cooperative:compression} details the development of new coding scheme and the establishment of bounds/capacity results in this work. Section~\ref{sec:discussions} provides further discussions and comparisons with some recently developed schemes. Finally, concluding remarks are given in Section~\ref{sec:conclusions}.

\par \textit{Notation}: For a pair of integers $z_1 \le z_2$, we use notation $\left[z_1:z_2\right]$ to denote the discrete interval $\{z_1,z_1+1,\cdots,z_2\}$. More generally, for any real number $c \ge 0$ and an integer $z_1 \le 2^c$, {we define $\left[z_1:2^c\right] = \{z_1,z_1+1,\cdots,2^{\lceil c \rceil}\}$ and $\left[z_1:2^c\right) = \{z_1,z_1+1,\cdots,2^{\lfloor c \rfloor}\}$ , where $\lceil.\rceil$ and $\lfloor.\rfloor$ is the ceiling function and the floor function, respectively}. Notation $\prod_{l \in {\mathcal A}}{\mathbb M}_l$ is used to denote the Cartesian product of sets $\{{\mathbb M}_l, l\in {\mathcal A}\}$. Notation $\left|{\mathcal A}\right|$ denotes the cardinality of a set~${\mathcal A}$. Notation ${\mathbb R}^{n}_{+}$ denotes the set of nonnegative real vectors in $n$ dimensions. {Notation $\text{conv}({\mathcal B})$ denotes the convex hull of a set ${\mathcal B}$, while notation $\overline {\mathcal B}$ denotes the closure of a set ${\mathcal B}$}. Finally, all the sets used in the paper are ordered sets.

\section{Problem Setup and Definitions}\label{sec:model}
\begin{figure*}[t]
\centering
\includegraphics[width=0.85\textwidth]{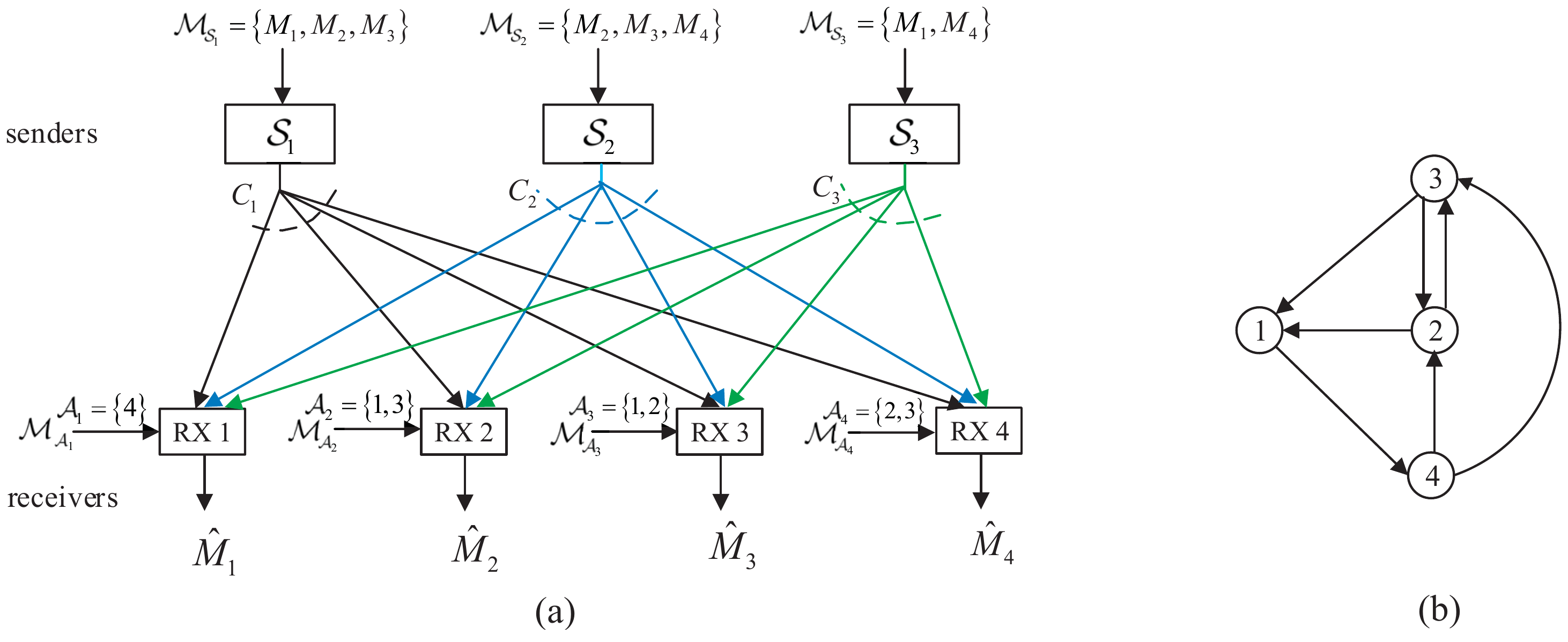}
\caption{(a) An example, with $K=3$ and $N=4$, of the multi-sender unicast index-coding problem where each sender is limited to know a subset of the messages; (b) the receiver side-information graph for the example in (a).}\label{fig:model}
\end{figure*}

We consider a multi-sender unicast index-coding problem that consists of the following:
\begin{itemize}
  \item $N$ independent messages, the collection of which is denoted by ${\mathcal M}=\left\{M_1,M_2,\cdots,M_N\right\}$;
  \item $K$ senders: each sender is indexed by a scalar $k$, $k\in[1:K]$, and the $k$-th sender having message indices ${\mathcal S}_k \subseteq [1:N]$ is also referred to as sender~${\mathcal S}_k$. These two sender descriptions are used interchangeably in the paper, and each is useful in different derivations. Then the messages available at sender ${\mathcal S}_k$ can now be represented by ${\mathcal M}_{{\mathcal S}_k} = \{M_j, j\in {\mathcal S}_k\} \subseteq {\mathcal M}$. Without loss of generality, assume that ${\mathcal M}_{{\mathcal S}_{k_1}} \neq {\mathcal M}_{{\mathcal S}_{k_2}}, \forall k_1 \neq k_2$, and $\cup _{k= 1}^K {\mathcal M}_{{\mathcal S}_k} = {\mathcal M}$;
  \item $N$ receivers: each receiver $j$ ($j\in \left[1:N\right]$) knows messages ${\mathcal M}_{{\mathcal A}_j}$ \textit{a priori}, i.e., a subset of ${\mathcal M}$ indexed by ${\mathcal A}_j \subseteq \left[1:N\right] \backslash \{j\}$, and requests message $M_j$;
  \item $K$ broadcast links: Each sender ${\mathcal S}_k$ is connected to all receivers via a noiseless broadcast link of an arbitrary link capacity $C_k > 0$ in bits/channel use (bcu).
\end{itemize}
Note that each sender contains a distinct subset of the messages ${\mathcal M}$. Therefore, with $N$ messages, the maximum number of admissible senders is $K_{\textrm{max}} = 2^{N}-1$, and thus we have $1 \le K \le K_{\textrm{max}}$. Fig.~\ref{fig:model}(a) depicts an example of the multi-sender index-coding problem with $K=3$ and $N=4$. It is also noted that Sadeghi~\textit{et~al.}~\cite{sadeghi2016distributed} considered a model with $K= K_{\textrm{max}}$ but allowed link capacity $C_k = 0$, i.e., $K_{\textrm{max}}$ senders are all present but some are inactive. So the problem described above and the one studied by Sadeghi~\textit{et~al.} {are equivalent}.

\par Given a sender setting, similarly to the single-sender setup, a multi-sender index-coding problem can be described by a receiver side-information digraph, $G$, with $N$ vertices, in which each vertex represents a receiver and an arc exists from vertex $i$ to vertex $j$ if and only if receiver $i$ has message $M_j$ (requested by receiver $j$) as its side information. Fig.~\ref{fig:model}(b) depicts the side-information digraph for the index-coding instance in Fig.~\ref{fig:model}(a).

\par We now define a multi-sender index code for the setup above:
\begin{definition}[Multi-Sender Index Code]\label{def:index:code}
Assume that each message $M_j$ is independently and uniformly distributed over the set ${\mathbb M}_j = \left[1:2^{nR_j}\right]$, where $n$ denotes the code block length and $R_j$ denotes the information bits per transmission, $j \in \left[1:N\right]$. A $\left(2^{nR_1},\cdots,2^{nR_N}, {2^{nC_1},\cdots,2^{nC_K}}, n\right)$ multi-sender index code consists of
\begin{enumerate}
  \item an encoder mapping $f_k$ at each sender ${\mathcal S}_k$:
  \begin{align}
  \prod_{i \in {\mathcal S}_k} {\mathbb M}_i \to {\left[1:2^{nC_k}\right)}, \label{equ:def:encodings}
  \end{align}
  which maps its messages to an index {$L_k \in \left[1:2^{nC_k}\right)$} sent to all receivers;
 \item and a decoder mapping $g_j$ at each receiver $j$:
  \begin{align}
  {\left[1:2^{nC_1}\right) \times \cdots \times \left[1:2^{nC_K}\right)} \times \prod_{i \in {\mathcal A}_j} {\mathbb M}_i \to {\mathbb M}_j, \label{equ:def:decodings}
  \end{align}
  which maps {its received indices $\{L_k,k\in[1:K]\}$ and its message side information ${\mathcal M}_{{\mathcal A}_j}$} to a message estimate ${\hat M}_j \in {\mathbb M}_j$.
\end{enumerate}
\end{definition}
\par The average probability of error is defined as $P_e^{(n)} = \Pr[({\hat M}_1,\cdots, {\hat M}_N) \ne \left({M}_1,\cdots, {M}_N\right)]$. A rate tuple $(R_1,\cdots,R_N)$ is said to be achievable {for the index-coding problem with a given link-capacity tuple $(C_1,\cdots, C_K)$, if there exists a sequence of $\left(2^{nR_1},\cdots,2^{nR_N}, 2^{nC_1},\cdots,2^{nC_K}, n\right)$} multi-sender index codes defined as above such that $P_e^{(n)} \to 0$ as $n \to \infty$. The capacity region~${\mathcal C}$ is the closure of the set of achievable rate tuples. \par The goal is to characterize the full capacity region or to establish bounds on the capacity region of this multi-sender index-coding problem.

\section{Preliminaries}\label{sec:preliminaries}
\par For comparison, we revisit the key ideas and re-state related results by Sadeghi~\textit{et al.}~\cite[Corollary~2]{sadeghi2016distributed} using our notation for consistency.

\begin{proposition}[{An Outer Bound}]\label{corollary:MAIS}
Given any fixed sender configuration and link capacities $\{C_k, k\in[1:K]\}$, and for a multi-sender index-coding problem represented by the side-information digraph $G$, if a rate tuple $(R_1,\cdots,R_N)$ is achievable, it must satisfy
\begin{align} \label{equ:MAIS:outerbound}
\sum\limits_{j \in S} {R_j } \le \sum\limits_{k \in [1:K] :~{\mathcal S}_k  \cap S \ne \emptyset} {C_k},
\end{align}
for all $S \subseteq \left[1:N\right]$ for which the subgraph of $G$ induced by $S$ is acyclic.
\end{proposition}

\begin{remark}
{This outer bound can be interpreted as a generalized version of the maximal-acyclic-induced subgraph (MAIS) bound from the single-sender problem, and it {was expressed differently by Sadeghi~\textit{et al.}~\cite{sadeghi2016distributed}.} In Appendix~\ref{appendix:MAIS}, we provide an alternative proof of this expression. \remarkend}
\end{remark}

\begin{figure*}[t]
\centering
\includegraphics[width=1.0\textwidth]{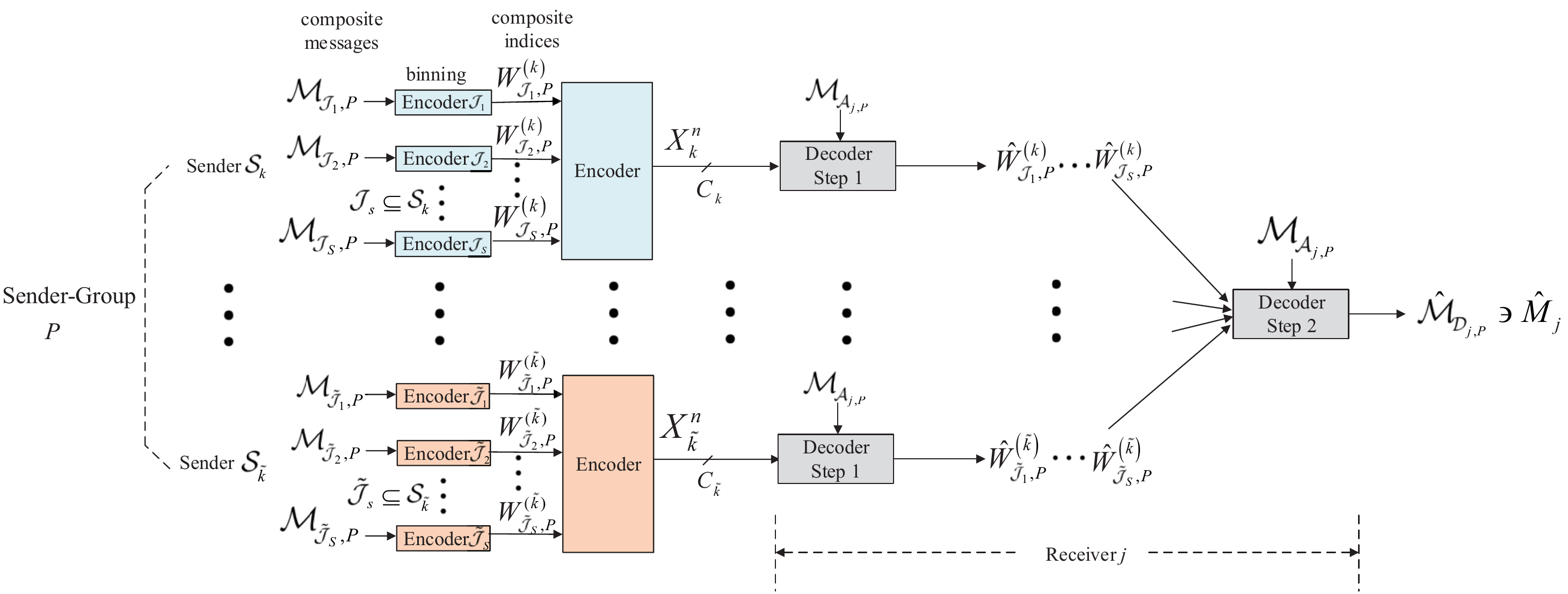}
\caption{The encoding operations at senders and decoding operations at any receiver $j \in {\mathcal S}_P$ in partitioned DCC.}\label{fig:DCC:encdec}
\end{figure*}

\par To define the inner bound, we use $\Pi$ to denote a partition of the set of all senders into disjoint subsets, and $P$ to denote a subset (a sender-group) in $\Pi$. {Therefore, sender groups in any partition $\Pi$ must satisfy}: \textit{i})~each $P \subseteq [1:K]$, s.t. $P \neq \emptyset$ and $\bigcup\nolimits_{P \in \Pi} P  = \left[ {1:K} \right]$; and~\textit{ii})~$P_1 \cap P_2 = \emptyset$, for any $P_1,P_2 \in \Pi$. If $\left|\Pi\right| = 1$, all senders are in the same group and this is just a trivial partition. Let ${\mathbf{\Pi}}_{\text{S}}$ denote the collection of all possible partitions of $K$ senders. In addition, for a given partition $\Pi \in {\mathbf{\Pi}}_{\text{S}}$ and for any sender-group $P \in \Pi$, we use ${\mathcal S}_P$ to denote the union of message indices available at the senders in $P$, i.e., ${\mathcal S}_P =\bigcup\nolimits_{k \in P} {{\mathcal S}_k}$. For any $j \in {\mathcal S}_P$, let ${\mathcal A}_{j,P} = {\mathcal A}_{j}\cap {\mathcal S}_P$ denote the side information receiver $j$ has in $P$, and let ${\mathcal D}_{j,P} \subseteq {\mathcal S}_P {\backslash {\mathcal A}_{j,P}}$, s.t.~$j \in {\mathcal D}_{j,P}$, denote the index of messages that receiver $j$ chooses to decode from the senders in $P$. Note that by having a receiver $j$ decode more messages than it requires, {that is, $\{j\} \subsetneq {\mathcal D}_{j,P}$, we may increase its message rate} by decoding and canceling some ``interfering" messages. The following proposition re-states an inner bound attained by partitioned DCC.

\begin{proposition}[{{\cite[Section~IV.C]{sadeghi2016distributed}}}]\label{proposition:DCC}
{Consider an arbitrary sender partition $\Pi$. The following rate region ${\mathcal R}_{\text{S}}(\Pi)$ is achievable by partitioned DCC:
\begin{align}
{\mathcal R}_{\text{S}}(\Pi) \buildrel \Delta \over = \left\{ \begin{array}{l}
 \left( {R_1 ,R_2 , \cdots ,R_N } \right) \in {\mathbb R}_+^N: \\
R_j =  \sum\limits_{P\in\Pi:~j \in {\mathcal S}_P} {R_{j,P} },~~j \in \left[ {1:N} \right] \\
\text{s.t.}~~({R_{j,P} ,j\in {\mathcal S}_P }) \in {\mathcal R}_P, \forall P \in \Pi
 \end{array} \right\}, \label{equ:DCC:FM1}
\end{align}
where ${\mathcal R}_P$ is the union of multiple polymatroidal rate regions under all possible decoding choices for the receivers in sender-group $P$:
\begin{align}\label{equ:DCC:union}
&{\mathcal R}_P = \nonumber\\
&\bigcup\limits_{\left\{ {\scriptstyle {\mathcal D}_{j,P}  \subseteq {\mathcal S}_P{\backslash {\mathcal A}_{j,P}}: \hfill \atop
  \scriptstyle ~j \in {\mathcal D}_{j,P},\forall j \in {\mathcal S}_P  \hfill} \right\}} {{\mathcal R}\left({\left\{ {{\mathcal D}_{j,P} ,j \in {\mathcal S}_P } \right\}\left| {\left\{ {{\mathcal A}_{j,P} ,j \in {\mathcal S}_P } \right\}} \right.}\right)},
\end{align}
with rate region ${{\mathcal R}\left({\left\{ {{\mathcal D}_{j,P} ,j \in {\mathcal S}_P } \right\}\left| {\left\{ {{\mathcal A}_{j,P} ,j \in {\mathcal S}_P } \right\}} \right.}\right)}$ under any fixed decoding choice $\{{\mathcal D}_{j,P} ,j \in {\mathcal S}_P \}$ given by
\begin{align}\label{equ:DCC:mainproblem}
{\mathcal R}&\left({\left\{ {{\mathcal D}_{j,P} ,j \in {\mathcal S}_P } \right\}\left| {\left\{ {{\mathcal A}_{j,P} ,j \in {\mathcal S}_P } \right\}} \right.}\right)= \nonumber \\
&\left\{ \begin{array}{l}
 ({R_{j,P} ,j\in {\mathcal S}_P })  \in {\mathbb R}^{\left| {{\mathcal S}_P } \right|}_+ : \\
 \text{(a):}~\sum\limits_{i \in {\mathcal T}_j } {R_{i,P} }~{<}~\sum\limits_{\scriptstyle {\mathcal J}_1  \subseteq {\mathcal D}_{j,P}  \cup {\mathcal A}_{j,P} : \hfill \atop
  ~~\scriptstyle {\mathcal J}_1  \cap {\mathcal T}_j  \ne \emptyset  \hfill} ~{\sum\limits_{\scriptstyle k:k \in P, \hfill \atop
  \scriptstyle {\mathcal J}_1  \subseteq {\mathcal S}_k  \hfill} {\gamma _{{\mathcal J}_1,P}^{(k)} } },\\
 ~~~~~~~~~~~~~~~\forall {\mathcal T}_j \subseteq {{\mathcal D}_{j,P}},~\forall j \in {\mathcal S}_P, \\
 \text{s.t.} \\
 \text{(b):}~\sum\limits_{{\mathcal J}_2  \subseteq {\mathcal S}_k: {\mathcal J}_2  \not\subseteq {\mathcal A}_{j,P} } {\gamma _{{\mathcal J}_2,P}^{(k)} }~{<}~C_k,\\
 ~~~~~~~~~~~~~~~~~~~~~~~~~~\forall j \in {\mathcal S}_P,~\forall k \in P;\\
 ~~~~~~~~~~~{\gamma _{{\mathcal J},P}^{(k)}} \ge  0,~~\forall {\mathcal J} \subseteq {\mathcal S}_k, \forall k \in P.
 \end{array} \right\}.
\end{align}}
\end{proposition}

{\begin{corollary}
The general achievable rate region ${\mathcal R}_{\text{S}}$ attained by partitioned DCC is given by the {convex hull} of the union of all possible ${\mathcal R}_{\text{S}}(\Pi)$'s:
\begin{align}
{\mathcal R}_{\text{S}} = {\text{conv}\bigcup\limits_{\Pi \in {\mathbf{\Pi}}_{\text{S}}} {{\mathcal R}_{\text{S}}(\Pi)}}. \label{equ:convexhull:RS}
\end{align}
\end{corollary}}

\begin{remark}\label{remark:time:sharing}
The \textit{convex hull} operation in~\eqref{equ:convexhull:RS} is used to convexify the achievable rate region and can be beneficial in enlarging the region. It can be viewed as the incorporation of a \textit{time-sharing} strategy~\cite{YKGamalBook} among different sender partitions $\{\Pi \in {\mathbf{\Pi}}_{\text{S}} \}$.
\par In particular, consider that all senders are restricted to be in the same group, i.e., $\Pi = \{[1:K]\}$. The achievable rate region in this special case is then given by: ${\mathcal R} = {\text{conv} ({\mathcal R}_{\text{S}}(\{[1:K]\}))}$, where the convex union (time-sharing) {operates over} all decoding choices in~\eqref{equ:DCC:union}.~\remarkend
\end{remark}

\begin{remark}
The basic idea of partitioned DCC consists of partitioning senders into non-overlapping sender-groups, {splitting any message that appears in different sender-groups, solving the composite coding problem for each group, and then combining the corresponding sub-message rates}.

\par Specifically, with sender partition $\Pi$, each message $M_j \in [1:2^{nR_j}]$ is split as $M_j=(M_{j,P})_{{P\in \Pi: j\in {\mathcal S}_P}}$ s.t. $M_{j,P} \in [1:2^{nR_{j,P}}]$ and $\sum\nolimits_{P\in\Pi:~j \in {\mathcal S}_P } {R_{j,P}} = R_j,~j \in \left[ {1:N} \right]$. Finding the rate region of~\eqref{equ:DCC:mainproblem} corresponds to solving the composite coding problem for a group $P$ under a fixed decoding choice.

\par {In particular, in group $P$, let composite message at sender ${\mathcal S}_k$ refer to a non-empty {subtuple} of its messages $({\mathcal M}_{j,P}: j \in {\mathcal S}_k)$. Then sender ${\mathcal S}_k$ has a set of composite messages $\{{\mathcal M}_{\mathcal J,P}: {\mathcal J}\subseteq {\mathcal S}_k\}$, where ${\mathcal M}_{\mathcal J,P} \buildrel \Delta \over = (M_{j,P}: j \in {\mathcal J})$}. As illustrated in Fig.~\ref{fig:DCC:encdec}, sender ${\mathcal S}_k$ independently uses point-to-point compression (binning) to compress each ${\mathcal M}_{\mathcal J,P}$ to ${W}_{\mathcal J,P}^{(k)} \in {[1:2^{n\gamma_{{\mathcal J},P}^{(k)}})}$. Here, ${W}_{\mathcal J,P}^{(k)}$ is a composite index and $\gamma_{{\mathcal J},P}^{(k)} \ge 0$ is referred to composite rate. All composite indices at ${\mathcal S}_k$ are then encoded and conveyed to receivers by a flat coding $X^n_k({W}_{\mathcal J,P}^{(k)}, {\mathcal J}\subseteq {\mathcal S}_k)$.
\par Upon reception, in the first step, each receiver $j$ first recovers each sender's composite indices separately based on its side-information, which hence leads to $\left|P\right|$ constraints on the composite rates (see~\eqref{equ:DCC:mainproblem}.(b).) by each receiver. In the second step, also based on its side-information, each receiver employs simultaneous nonunique decoding~\cite{arbabjolfaei2013capacity} to decode a set of messages ${\mathcal M}_{{\mathcal D}_{j,P}}$ through a set of relevant composite indices. The resultant achievable rates are bounded by~\eqref{equ:DCC:mainproblem}.(a). In this way, the rate region under a fixed decoding choice can be obtained by eliminating all composite rate variables in~\eqref{equ:DCC:mainproblem}, e.g., by Fourier-Motzkin elimination~\cite[Appendix~D]{YKGamalBook}.
\par The overall achievable rate region ${\mathcal R}_P$ for each sender group $P$ is the union of rate regions evaluated for all possible decoding choices at the receivers involved (see~\eqref{equ:DCC:union}). A combined rate region ${\mathcal R}_{\text S}(\Pi)$ under partition $\Pi$ is then obtained via~\eqref{equ:DCC:FM1} by eliminating variables $\left\{ {R_{j,P} ,j \in {\mathcal S}_P, P \in \Pi} \right\}$ through Fourier-Motzkin elimination. Finally, the general achievable rate region ${\mathcal R}_{\text{S}}$ is given by the {convex hull} of the union of all possible ${\mathcal R}_{\text{S}}(\Pi)$'s as defined in~\eqref{equ:convexhull:RS}.~\remarkend
\end{remark}

\begin{remark}
It was indicated~\cite{sadeghi2016distributed} that ${\mathcal R}$ is optimal for all non-isomorphic index-coding instances with $N=3$ receivers and with arbitrary link capacities for senders. In addition, it was shown that ${\mathcal R}_{\text{S}}(\Pi)$ with appropriate sender partition $\Pi$ strictly improves upon ${\mathcal R}$ for an instance with $N=4$ receivers and $K=15$ senders each with unit link capacity. However, partitioned DCC is still suboptimal in general.~\remarkend
\end{remark}

\section{{Main Results}}\label{sec:cooperative:compression}
\par In this section, we propose a new coding scheme that provides two conceptual improvements over partitioned DCC. Subsection~\ref{sub:sec:inner:bound} presents the general scheme and establishes the corresponding inner bound. Subsections~\ref{sub:sec:benefit:cooperative} and \ref{sub:sec:benefit:link} then elaborate the benefit of each conceptual improvement via several index-coding instances.

\subsection{General Achievable Scheme and Inner Bound}\label{sub:sec:inner:bound}

\par The improvements are motivated by the following two limitations in partitioned DCC:
\begin{itemize}
  \item Within each sender-group, a sender always treats its composite messages as independent source data and employs point-to-point compression (binning) to generate composite indices to broadcast to receivers, regardless of whether or not some messages might be common to two or more senders in the group. This strategy consequently fails to exploit the potential overlapping of messages at different senders for more efficient common description of composite messages;
  \item Each sender belongs exclusively to a sender-group and exhausts its link capacity for the transmission in that group. This strategy can be sub-optimal in some instances, as it precludes senders carefully allocating their resources and contributing to collaborative transmissions in different groups.
\end{itemize}

\par To overcome the first limitation, we propose a \textit{cooperative compression} technique, which encodes the same composite message to the same composite index at different senders. Each sender then will have a mixed set of private and common composite indices. In this way, the senders can collaboratively describe common composite indices. We will show that, this cooperative compression is more effective than point-to-point compression in the sense that it can support a larger composite rate region, which in turn leads to a larger message rate region.

\par To overcome the second limitation, we propose a \textit{joint link-and-sender partitioning} technique, in which each sender is allowed to split its link capacity appropriately so as to participate in multiple link-sender groups if necessary. This joint partitioning technique hence contains the existing sender partitioning as a special case.

\par Based on these two techniques, we now propose a new achievable scheme for the multi-sender index-coding problem. The scheme in its most general form consists of forming different link-sender groups by joint link-and-sender partitioning, splitting any message that appears in different groups, implementing composite coding with cooperative compression of composite messages, solving the composite coding problem for each group and then combining the corresponding sub-message rates. This scheme is termed as multi-sender \textit{Cooperative Composite Coding} (CCC).

\par To characterize the inner bound attained by CCC, we use $({\tilde \Pi}, {\mathbf C}_{\tilde \Pi})$ to denote a joint link-and-sender partition, where ${\tilde \Pi}$ is a set of sender subsets $\{\tilde P\}$\footnote{{In contrast to $\Pi$, ${\tilde \Pi}$ is not a partition of the senders.}}, and ${\mathbf C}_{\tilde \Pi} = [C_{k,\tilde P}: \forall k\in {\tilde P}, \forall {\tilde P} \in {\tilde \Pi}]$ is a link-partitioning vector for a given ${\tilde \Pi}$. We further define $({\tilde P}, {\mathbf C}_{\tilde P})$ to represent a link-sender group, where ${\tilde P} \in {\tilde \Pi}$ {is the set of sender indices in the group} and ${\mathbf C}_{\tilde P} = [C_{k,\tilde P}: \forall k\in {\tilde P}]$ {are the link fractions allocated to sender group ${\tilde P}$}. Note that ${\tilde \Pi}$ is said to be admissible if: \textit{i})~each ${\tilde P} \subseteq [1:K]$, s.t. ${\tilde P} \neq \emptyset$ and $\bigcup\nolimits_{{\tilde P} \in {\tilde \Pi}} {\tilde P}  = \left[ {1:K} \right]$; and~\textit{ii})~${\tilde P}_1 \neq {\tilde P}_2$, for any ${\tilde P}_1,{\tilde P}_2 \in {\tilde \Pi}$. {Let ${\mathbf{\tilde \Pi}}_{\text{LS}}$ denote the set of all admissible ${\tilde \Pi}$'s. For any ${\tilde \Pi} \in {\mathbf{\tilde \Pi}}_{\text{LS}}$}, ${\mathbf C}_{\tilde \Pi}$ is said to be admissible if $\sum\nolimits_{\tilde P \in \tilde \Pi: k \in \tilde P} {C_{k,\tilde P}} \le C_k$, with each ${C_{k,\tilde P}}>0, \forall k \in {\tilde P}, \forall {\tilde P} \in {\tilde \Pi}$.

\par We also use notations consistent with sender-partitioning in partitioned DCC. {In particular, given any link-sender group $({\tilde P}, {\mathbf C}_{\tilde P})$, let ${\mathcal S}_{\tilde P} = \bigcup\nolimits_{k \in {\tilde P}}{\mathcal S}_k$ be the union of message indices at the senders in ${\tilde P}$, let ${\mathcal A}_{j,{\tilde P}} = {\mathcal A}_{j}\cap {{\mathcal S}_{\tilde P}}$ be the side information receiver $j$ has in ${\tilde P}$, and let ${\mathcal D}_{j,{\tilde P}} \subseteq {\mathcal S}_{\tilde P}{\backslash {\mathcal A}_{j,{\tilde P}}}$, s.t.~$j \in {\mathcal D}_{j,{\tilde P}}$, be the indices of messages that receiver $j$ decodes from the senders in~${\tilde P}$}.

\begin{figure*}[t]
\centering
\includegraphics[width=1.0\textwidth]{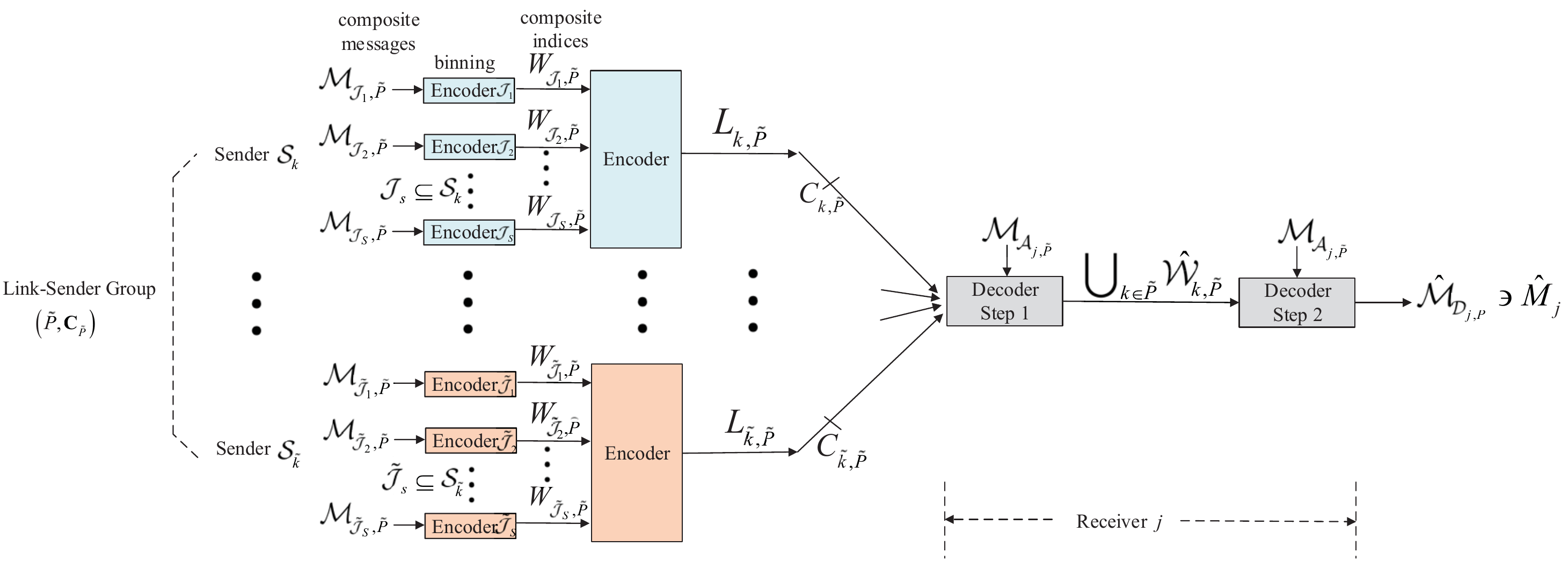}
\caption{The encoding operations at senders within $\tilde P$ and decoding operations at any receiver $j \in {\mathcal S}_{\tilde P}$ in the proposed CCC.}\label{fig:CCC:encdec}
\end{figure*}

\par The following proposition states the new inner bound attained by CCC.
\begin{proposition}[CCC Rate Region]\label{proposition:MCCC}
{Consider arbitrary admissible ${\tilde \Pi} \in {\mathbf{\tilde \Pi}}_{\text{LS}}$ as defined}. The following rate region {${\mathcal R}_{\text{CLS}} (\tilde \Pi, {\mathbf C}_{\tilde \Pi})$} is achievable by CCC:
\begin{align}
&{\mathcal R}_{\text{CLS}} (\tilde \Pi, {\mathbf C}_{\tilde \Pi}) \buildrel \Delta \over = \nonumber\\
&\left\{ \begin{array}{l}
 \left( {R_1 ,R_2 , \cdots ,R_N } \right) \in {\mathbb R}_+^N: \\
R_j = \sum\limits_{{\tilde P} \in {\tilde \Pi}: j \in {\mathcal S}_{\tilde P}} {R_{j,\tilde P} },~~j \in \left[ {1:N} \right] \\
\text{s.t.}~~(R_{j,{\tilde P}}, j \in {{\mathcal S}_{\tilde P}}) \in {\mathcal R}_{(\tilde P,{\mathbf C}_{\tilde P})},\forall {\tilde P} \in {\tilde \Pi},\\
~~~~~~{\text{for some admissible~}{\mathbf C}_{\tilde \Pi}}. \\
 \end{array} \right\}, \label{equ:MCCC:FM1}
\end{align}
where ${\mathcal R}_{(\tilde P,{\mathbf C}_{\tilde P})}$ is the union of multiple polymatroidal rate regions under all possible decoding choices for the receivers in link-sender group ${(\tilde P,{\mathbf C}_{\tilde P})}$:
\begin{align}\label{equ:MCCC}
&{\mathcal R}_{(\tilde P,{\mathbf C}_{\tilde P})} = \nonumber \\
&\bigcup\limits_{\left\{ {\scriptstyle {\mathcal D}_{j,{\tilde P}} \subseteq {\mathcal S}_{{\tilde P}}{\backslash {\mathcal A}_{j,{\tilde P}}}: \hfill \atop
  \scriptstyle ~j \in {\mathcal D}_{j,{\tilde P}}, \forall j \in {\mathcal S}_{{\tilde P}}  \hfill} \right\}} {{\mathcal R}({\{ {{\mathcal D}_{j,{\tilde P}},j \in {\mathcal S}_{\tilde P} }\},{\mathbf C}_{\tilde P}| {\{ {{\mathcal A}_{j,{\tilde P}} ,j \in {\mathcal S}_{\tilde P}}\}}})},
\end{align}
with rate region ${{\mathcal R}({\{ {{\mathcal D}_{j,{\tilde P}},j \in {\mathcal S}_{\tilde P} }\},{\mathbf C}_{\tilde P}| {\{ {{\mathcal A}_{j,{\tilde P}},j \in {\mathcal S}_{\tilde P}}\}}})}$ under any fixed decoding choice $\{{\mathcal D}_{j,{\tilde P}} ,j \in {\mathcal S}_{\tilde P} \}$ given by:
\begin{align}
&{{\mathcal R}({\{ {{\mathcal D}_{j,{\tilde P}},j \in {\mathcal S}_{\tilde P} }\},{\mathbf C}_{\tilde P}| {\{ {{\mathcal A}_{j,{\tilde P}},j \in {\mathcal S}_{\tilde P}}\}}})} =  \nonumber\\
&~~~~~~\left\{ \begin{array}{l}
 (R_{j,{\tilde P}}, j \in {{\mathcal S}_{\tilde P}}) \in {\mathbb R}^{\left| {{\mathcal S}_{\tilde P} } \right|}_+ : \\
 \text{(a):}~\sum\limits_{i \in {\mathcal T}_j } { R_{i,{\tilde P}} }~{<}~\sum\limits_{\scriptstyle {\mathcal J}_1  \subseteq {\mathcal D}_{j,{\tilde P}}  \cup {\mathcal A}_{j,{\tilde P}} : \hfill \atop
  ~~\scriptstyle {\mathcal J}_1  \cap {\mathcal T}_j  \ne \emptyset  \hfill} ~{{\gamma_{{\mathcal J}_1, \tilde P}} },\\
 ~~~~~~~~~~~~~~~\forall {\mathcal T}_j \subseteq {\mathcal D}_{j,{\tilde P}},~\forall j \in {\mathcal S}_{\tilde P}, \\
 \text{s.t.} \\
 \text{(b):}~\sum\limits_{\scriptstyle ~~~~~{\mathcal J}_2 \in {\mathcal I}_{{\tilde {\mathcal K}}}: \hfill \atop \scriptstyle {\mathcal J}_2 \notin {\mathcal I}_{{\tilde {\mathcal K}}^c},~{\mathcal J}_2 \not\subseteq {\mathcal A}_{j,{\tilde P}} } {\gamma _{{\mathcal J}_2, {\tilde P}} }~{<}~\sum\limits_{k \in \tilde{\mathcal K}} C_{k,{\tilde P}},\\
 ~~~~~~~~~~~~~~~~~~~~~~~~~~\forall j \in {\mathcal S}_{\tilde P},~\forall {\tilde{\mathcal K}} \subseteq {\tilde P}; \\
 ~~~~~~~~~~~~~~~~{\gamma_{{\mathcal J},{\tilde P}}}\ge 0,\forall {\mathcal J} \subseteq {\mathcal S}_k, \forall k \in {\tilde P}.
 \end{array} \right\}, \label{equ:MCCC:rate}
\end{align}
in which for any ${\tilde {\mathcal K}} \subseteq {\tilde P}$, let ${\tilde {\mathcal K}}^c = {\tilde P}\backslash{\tilde {\mathcal K}}$, and ${\mathcal I}_{{\tilde {\mathcal K}}} = \bigcup\nolimits_{k\in {\tilde {\mathcal K}}}\{{\mathcal J}: {\mathcal J} \subseteq {\mathcal S}_k\}$.
\par The general achievable rate region ${\mathcal R}_{\text{CLS}}$ attained by CCC is given by the {convex hull} of the union of all possible ${\mathcal R}_{\text{CLS}}(\tilde \Pi, {\mathbf C}_{\tilde \Pi})$'s:
\begin{align}
{\mathcal R}_{\text{CLS}} = {\text{conv}\bigcup\limits_{{\tilde \Pi} \in {\mathbf{\tilde \Pi}}_{\text{LS}}} {{\mathcal R}_{\text{CLS}} (\tilde \Pi, {\mathbf C}_{\tilde \Pi})}}. \label{equ:convexhull:RCLS}
\end{align}
\end{proposition}

\begin{IEEEproof}
{Consider any ${\tilde \Pi}$ as given. Each message $M_j \in [1:2^{nR_j}]$ is split as $M_j=(M_{j,{\tilde P}})_{{{\tilde P}\in \tilde \Pi: j\in {\mathcal S}_{\tilde P}}}$ s.t. $M_{j,{\tilde P}} \in [1:2^{nR_{j,{\tilde P}}}]$ and $\sum\nolimits_{{\tilde P}\in \tilde \Pi:~j \in {\mathcal S}_{\tilde P} } {R_{j,{\tilde P}}} = R_j,~j \in \left[ {1:N} \right]$. Consider arbitrary link-sender group ${(\tilde P,{\mathbf C}_{\tilde P})}$ in the joint link-and-sender partition}. We first prove the achievability of~\eqref{equ:MCCC:rate} under any fixed decoding choice $\{{\mathcal D}_{j,{\tilde P}},j \in {\mathcal S}_{\tilde P}\}$.

\par The encoding operations at each sender and the decoding operations at any receiver~$j$ are illustrated in Fig.~\ref{fig:CCC:encdec}. Specifically, each sender ${\mathcal S}_k$ has a set of composite messages $\{{\mathcal M}_{{\mathcal J},{\tilde P}}: {\mathcal J}\subseteq {\mathcal S}_k\}$, where ${\mathcal M}_{\mathcal J,{\tilde P}} \buildrel \Delta \over = {(M_{j,{\tilde P}}: j \in {\mathcal J})}$. For the encoding, in the first step, each sender ${\mathcal S}_k$ encodes (compresses) {each composite message ${\mathcal M}_{\mathcal J,{\tilde P}}$} into a composite index $W_{{\mathcal J}, \tilde P} \in {[1:2^{n\gamma_{{\mathcal J}, \tilde P}})}$ at composite rate $\gamma_{{\mathcal J}, \tilde P} \ge 0$ bcu via a standard random binning~\cite{YKGamalBook}. Denote the collection of all composite indices at sender ${\mathcal S}_k$ by ${\mathcal W}_{k,{\tilde P}} = \{W_{{\mathcal J}, \tilde P}, {\mathcal J}\subseteq {\mathcal S}_k\}$. Whenever two or more senders have a composite message ${\mathcal M}_{{\mathcal J},{\tilde P}}$ in common, they will encode to the same composite index $W_{{\mathcal J}, \tilde P}$ (i.e., a cooperative compression strategy). Hence, ${\mathcal W}_{k,{\tilde P}}$ and ${\mathcal W}_{{\tilde k},{\tilde P}}$ can be correlated if senders ${\mathcal S}_k$ and ${\mathcal S}_{\tilde k}$ have some common message. In this way, in the next step, the description of composite indices from all senders to any receiver can now be viewed as a Slepian-Wolf-Cover like problem of transmitting multiple correlated sources through {a multiple-access channel with orthogonal links $\{C_{k,\tilde P}, k\in \tilde P\}$~\cite{slepian1973noiseless,cover1975proof,han1980slepian}}, and with some side-information at each receiver. Since source-channel separation is optimal for this class of multiple-access channels with orthogonal links~\cite{han1980slepian}, each sender ${\mathcal S}_k$ uses Slepian-Wolf-Cover random binning to encode ${\mathcal W}_{k,{\tilde P}}$ into a bin index $L_{k,\tilde P} \in {[1:2^{nC_{k,\tilde P}})}$ at rate $C_{k,\tilde P}$ bcu and then noiselessly broadcasts the index to receivers.

\par Upon reception, receiver $j$ ($j\in{\mathcal S}_{\tilde P}$) retrieves all bin indices~$\{L_{1,\tilde P},\cdots,L_{K,\tilde P}\}$ and attempts to recover all composite indices from the senders by leveraging its side-information. Receiver $j$ can reliably recover all composite indices within the group with vanishing probability of error as the number of channel uses $n\to \infty$, if the following conditions can be satisfied according to Han~\cite[Section~I]{han1980slepian} (also, see Cover~\cite[Theorem~2]{cover1975proof}):
\begin{align}
\frac{1}{n}H( {{{\mathcal W}}_{{\tilde{\mathcal K}},\tilde P} \left| {{{\mathcal W}}_{{\tilde{\mathcal K}}^c ,\tilde P} ,{{\mathcal M}}_{{{\mathcal A}}_{j,\tilde P} } } \right.})~<~\sum\limits_{k \in \tilde{\mathcal K}} {C_{k,\tilde P} },~\forall {\tilde{\mathcal K}} \subseteq {\tilde P},\label{equ:composite:rate:conditions}
\end{align}
where ${{\mathcal W}}_{{\tilde{\mathcal K}},\tilde P} = \bigcup\nolimits_{k \in \tilde{\mathcal K}}{{\mathcal W}}_{k,\tilde P}$ and ${\tilde{\mathcal K}^c}$ denotes the complement of $\tilde{\mathcal K}$ with respect to $\tilde P$, {and the extra conditioning on ${{\mathcal M}}_{{{\mathcal A}}_{j,\tilde P} }$ is due to the presence of message side information ${{\mathcal M}}_{{{\mathcal A}}_{j,\tilde P} }$ at receiver~$j$}. Recall the notation ${\mathcal I}_{{\tilde {\mathcal K}}} = \bigcup\nolimits_{k\in {\tilde {\mathcal K}}}\{{\mathcal J}: {\mathcal J} \subseteq {\mathcal S}_k\}$ we have defined. With this notation, {as $n\to \infty$, we have that}
\begin{align}
H( {{{\mathcal W}}_{{\tilde{\mathcal K}},\tilde P} \left| {{{\mathcal W}}_{{\tilde{\mathcal K}}^c ,\tilde P} ,{{\mathcal M}}_{{{\mathcal A}}_{j,\tilde P} } } \right.}) = n\sum\limits_{\scriptstyle ~~~~~{\mathcal J}_2 \in {\mathcal I}_{{\tilde {\mathcal K}}}: \hfill \atop \scriptstyle {\mathcal J}_2 \notin {\mathcal I}_{{\tilde {\mathcal K}}^c},~{\mathcal J}_2 \not\subseteq {\mathcal A}_{j,{\tilde P}} } {\gamma _{{\mathcal J}_2, {\tilde P}}}.
\end{align}
{Therefore, the conditions in~\eqref{equ:composite:rate:conditions} hold if}
\begin{align}
 {\sum\limits_{\scriptstyle ~~~~~{\mathcal J}_2 \in {\mathcal I}_{{\tilde {\mathcal K}}}: \hfill \atop \scriptstyle {\mathcal J}_2 \notin {\mathcal I}_{{\tilde {\mathcal K}}^c},~{\mathcal J}_2 \not\subseteq {\mathcal A}_{j,{\tilde P}} } {\gamma _{{\mathcal J}_2, {\tilde P}} }~<~\sum\limits_{k \in \tilde{\mathcal K}} C_{k,{\tilde P}},~\forall j \in {\mathcal S}_{\tilde P},~\forall {\tilde{\mathcal K}} \subseteq {\tilde P},} \label{equ:composite:rate:conditions:a}
\end{align}
which lead to constraints (b) of~\eqref{equ:MCCC:rate}. Each constraint reads that the sum of composite rates for composite indices (excluding those known at receiver $j$ by side information) that can only be generated by senders in ${\tilde{\mathcal K}}$ is bounded by the sum link capacity of these senders.

\par Given all composite indices recovered, receiver $j$ then chooses a proper decoding set ${\mathcal D}_{j,\tilde P}$ such that $j \in {\mathcal D}_{j,\tilde P}$ and employs simultaneous nonunique decoding~\cite{arbabjolfaei2013capacity} to decode messages ${\mathcal M}_{{\mathcal D}_{j,\tilde P}}$ by utilizing its side information. The messages can be successfully decoded with vanishing probability of error as the number of channel uses $n \to \infty$, if the following bounds are satisfied~\cite{arbabjolfaei2013capacity}:
\begin{align}
\sum\limits_{i \in {\mathcal T}_j } { R_{i,{\tilde P}} }~{<}~\sum\limits_{\scriptstyle {\mathcal J}_1  \subseteq {\mathcal D}_{j,{\tilde P}}  \cup {\mathcal A}_{j,{\tilde P}} : \hfill \atop
  ~~\scriptstyle {\mathcal J}_1  \cap {\mathcal T}_j  \ne \emptyset  \hfill} ~{{\gamma_{{\mathcal J}_1, \tilde P}} }, \label{equ:MCCC:rate:constraints2}\\
 ~~~~~~~~~~~~~~~\forall {\mathcal T}_j \subseteq {\mathcal D}_{j,{\tilde P}},~\forall j \in {\mathcal S}_{\tilde P}, \label{equ:MCCC:rate:constraints}
\end{align}
which lead to bounds (a) of~\eqref{equ:MCCC:rate}. This hence proves the achievability of~\eqref{equ:MCCC:rate} under a fixed decoding choice for link-sender group $(\tilde P,{\mathbf C}_{\tilde P})$, and the rate region can be characterized by eliminating all composite rate variables through Fourier-Motzkin elimination.

\par The overall rate region ${\mathcal R}_{(\tilde P,{\mathbf C}_{\tilde P})}$ is the union of rate regions evaluated for all possible decoding choices at the receivers involved (see~\eqref{equ:MCCC}). A combined rate region {${\mathcal R}_{\text{CLS}}(\tilde \Pi, {\mathbf C}_{\tilde \Pi})$} is then obtained by eliminating variables $\{ {R_{j,{\tilde P}}, \forall j \in {\mathcal S}_{\tilde P}, \forall {\tilde P} \in {\tilde \Pi}}\}$ and $\{C_{k,\tilde P}, \forall k \in {\tilde P}, \forall {\tilde P} \in {\tilde \Pi} \}$ via Fourier-Motzkin elimination (see~\eqref{equ:MCCC:FM1}). The final rate region ${\mathcal R}_{\text{CLS}}$ is obtained by taking the {convex hull} of the union of rate regions ${{\mathcal R}_{\text{CLS}}(\tilde \Pi, {\mathbf C}_{\tilde \Pi})}$'s under all possible $\tilde \Pi$'s (see~\eqref{equ:convexhull:RCLS}).
\end{IEEEproof}

\begin{remark}
By comparing encoding/decoding operations as shown in Fig.~\ref{fig:DCC:encdec} and Fig.~\ref{fig:CCC:encdec}, one can further understand the conceptual differences of partitioned DCC and CCC. When encoding, unlike partitioned DCC, CCC enables a collaborative description of composite indices from all senders within each group. When decoding, in partitioned DCC, each receiver decodes composite indices from each sender independently in the first step, while in CCC, each receiver decodes all composite indices within the group jointly in the first step. As a result, the composite rate constraints in CCC (see~\eqref{equ:MCCC:rate}.(b)) are less restrictive than those in partitioned DCC (see~\eqref{equ:DCC:mainproblem}.(b)), which will be shown in the proof of Proposition~\ref{proposition:CCC:over:DCC} in Appendix~\ref{appendix:proposition:CCC:over:DCC}.~\remarkend
\end{remark}

\begin{remark}\label{remark:CCC:special:cases}
The general CCC scheme contains the following two special cases: \textit{i})~with cooperative compression and with all senders in a group; and~\textit{ii})~with cooperative compression and sender-partitioning, whose corresponding achievable rate regions are denoted by ${\mathcal R}_{\text{C}}$ and ${\mathcal R}_{\text{CS}}$, respectively. In particular, ${\mathcal R}_{\text{C}} = {\text{conv}({{\mathcal R}_{\text{CLS}} ({\tilde \Pi},{\mathbf C}_{\tilde \Pi})})}$, where ${\tilde \Pi} = \{[1:K]\}$ and $C_{k,{\tilde P}={\tilde \Pi}} = C_k, \forall k\in[1:K]$, while {${\mathcal R}_{\text{CS}} = \text{conv}\bigcup\nolimits_{{\tilde \Pi} \in {\mathbf{\Pi}}_{\text{S}}} {{\mathcal R}_{\text{CLS}}({\tilde \Pi},{\mathbf C}_{\tilde \Pi})}$, where ${\tilde \Pi}$ is restricted to be a sender partition in ${\mathbf{\Pi}}_{\text{S}}$ and $C_{k,{\tilde P}} = C_k, \forall k\in {\tilde P}, \forall {\tilde P} \in {\tilde \Pi}$}. By definition, we thus have the following relationship
\begin{align}
 {\mathcal R}_{\text{C}} \subseteq {\mathcal R}_{\text{CS}} \subseteq {\mathcal R}_{\text{CLS}}, \label{equ:CCC:specicalcases}
 \end{align}
where both the inclusions can be strict as shown by examples in Subsection~\ref{sub:sec:benefit:link}. A special case of CCC suffices to achieve the capacity region for certain instances as shown later.~\remarkend
\end{remark}
\par Moreover, we can establish that CCC in general improves upon partitioned DCC.
\begin{proposition}\label{proposition:CCC:over:DCC}
{${\mathcal R}_{\text{S}} \subseteq {\mathcal R}_{\text{CLS}}$} in general, and the inclusion can be strict.
\end{proposition}
\par Proposition~\ref{proposition:CCC:over:DCC} can be established by proving that ${\mathcal R}_{\text{S}}\subseteq {\mathcal R}_{\text{CS}}$, as we already have ${\mathcal R}_{\text{CS}}\subseteq {\mathcal R}_{\text{CLS}}$ from~\eqref{equ:CCC:specicalcases}. The proof is given in Appendix~\ref{appendix:proposition:CCC:over:DCC}.

\subsection{The Benefit of Cooperative Compression}\label{sub:sec:benefit:cooperative}
We first examine a simple 2-sender index-coding example as defined below.
\begin{example}[{an instance where ${\mathcal R} \subsetneq {\mathcal R}_{\textup{C}}$ and $\overline{{\mathcal R}_{\textup{C}}}= {\mathcal C}$}]\label{example:DCC:fails}
In this example, there are
\begin{itemize}
  \item $K=2$ senders, with indices of messages ${\mathcal S}_1=\{1,2,3\}$, ${\mathcal S}_2=\{2,3,4\}$, and with link capacities $C_1 = C_2 = 1$~bcu;
  \item $N=4$ receivers, with side-information ${\mathcal A}_1 = \{4\}$, ${\mathcal A}_2 = \{1,3\}$, ${\mathcal A}_3 = \{1,2\}$, and ${\mathcal A}_4 = \{2,3\}$, respectively. This again corresponds to the same digraph $G$ as depicted in Fig.~\ref{fig:model}(b).
\end{itemize}
\end{example}
\par For this example, note that the MAIS outer bound of~\eqref{equ:MAIS:outerbound} is specialized to
\begin{align} \label{equ:eg2:outerbound}
{\mathcal R}_{\text{out}}^{\text{eg\ref{example:DCC:fails}}}  = \left\{ \begin{array}{l}
 \left( {R_1 ,R_2 ,R_3 ,R_4 } \right) \in {\mathbb R}^4_+  : \\
 R_1  \le 1,~~~~R_4  \le 1, \\
 R_1  + R_2  \le 2,~R_1  + R_3  \le 2, \\
 R_2  + R_4  \le 2,~R_3  + R_4  \le 2. \\
 \end{array} \right\}.
\end{align}
\par To derive the inner bound by partitioned DCC, we use Proposition~\ref{proposition:DCC} and observe that the best possible achievable rate region is attained by grouping these two senders together and time-sharing between the following two decoding choices (see \textit{Remark~\ref{remark:time:sharing}}):
\begin{enumerate}[(i)]
\item with ${\mathcal D}_{1} =\{1\}$ and ${\mathcal D}_{j} =[1:4]\backslash {\mathcal A}_j$~for~$j=2,3,4$, the following rate region is achieved:
\begin{align}
 {\mathcal R}_{1}^{\text{eg\ref{example:DCC:fails}}}  = \left\{ \begin{array}{l}
 \left( {R_1 ,R_2 ,R_3 ,R_4 } \right) \in {\mathbb R}^4_+: \\
 R_1  < 1,~~~~R_4  < 1, \\
 R_1  + R_3 + R_4  < 2, \\
 R_1  + R_2  + R_4  < 2. \\
 \end{array} \right\};
    \end{align}
\item with ${\mathcal D}_{j} =[1:4]\backslash {\mathcal A}_j$~for~$j=1,2,3,4$, the following rate region is achieved:
\begin{align}
 {\mathcal R}_{2}^{\text{eg\ref{example:DCC:fails}}}  = \left\{ \begin{array}{l}
 \left( {R_1 ,R_2 ,R_3 ,R_4 } \right) \in {\mathbb R}^4_+: \\
 R_1  < 1,~~~~R_4  < 1, \\
 R_3 + R_4 < 2,~R_2  + R_4  < 2,\\
 R_1 + R_2 + R_3 < 2. \\
 \end{array} \right\}.
    \end{align}
\end{enumerate}
Since none of ${\mathcal R}_{1}^{\text{eg\ref{example:DCC:fails}}}$ and ${\mathcal R}_{2}^{\text{eg\ref{example:DCC:fails}}}$ strictly contains the other, a {convex hull} of the union of these two regions yields the best possible achievable rate region ${\mathcal R}^{\text{eg\ref{example:DCC:fails}}}$ by partitioned DCC for the 2-sender index-coding problem considered:
\begin{align}
 {\mathcal R}^{\text{eg\ref{example:DCC:fails}}}  = \left\{ \begin{array}{l}
 \left( {R_1 ,R_2 ,R_3 ,R_4 } \right) \in {\mathbb R}^4_+: \\
 R_1  < 1,~~~~R_4  < 1, \\
R_1  + R_2  < 2,~R_1  + R_3  < 2,\\
R_2  + R_4  < 2,~R_3  + R_4  < 2,\\
~R_1 + R_2 + R_3 + 2R_4 < 4,\\
~2R_1 + R_2 + R_3 + R_4 < 4.
 \end{array} \right\}.
\end{align}
It can be seen that there is a gap between this inner bound and the MAIS outer bound~\eqref{equ:eg2:outerbound}. In particular, a symmetric-rate tuple $(R_j=1,j\in[1:4])$ is in ${\mathcal R}_{\text{out}}^{\text{eg\ref{example:DCC:fails}}}$, but it is not in {the closure of} ${\mathcal R}^{\text{eg\ref{example:DCC:fails}}}$ above.

\par We now explain why partitioned DCC is suboptimal here, and how CCC overcomes this shortcoming. In particular, consider the first decoding choice where ${\mathcal D}_{1} =\{1\}$ and ${\mathcal D}_{j} =[1:4]\backslash {\mathcal A}_j$~for~$j=2,3,4$ in the composite-coding formulation of~\eqref{equ:DCC:mainproblem}. To obtain~${\mathcal R}_{1}^{\text{eg\ref{example:DCC:fails}}}$, sender ${\mathcal S}_1$ compresses composite messages ${\mathcal M}_1$ and ${\mathcal M}_{2,3}$ into composite indices $W_1^{(1)}$ and ${W}_{2,3}^{(1)}$ at rates $\gamma_{1}^{(1)}$ and $\gamma_{2,3}^{(1)}$~bcu, respectively, and sets the rates of the remaining indices to zero, while sender ${\mathcal S}_2$ compresses ${\mathcal M}_4$ and ${\mathcal M}_{2,3}$ to composite indices $W_4^{(2)}$ and ${W}_{2,3}^{(2)}$ at rates $\gamma_{4}^{(2)}$ and $\gamma_{2,3}^{(2)}$~bcu, respectively, and sets the rates of the remaining indices to zero in~\eqref{equ:DCC:mainproblem}. {Then the constraints (redundant ones are discarded) in~\eqref{equ:DCC:mainproblem} read}
\begin{align}
&R_1 < \gamma_{1}^{(1)}, ~~R_2 < \gamma_{2,3}^{(1)} + \gamma_{2,3}^{(2)}, \\
&R_3 < \gamma_{2,3}^{(1)} + \gamma_{2,3}^{(2)}, ~R_4 < \gamma_{4}^{(2)},\\
&\gamma_1^{(1)} + \gamma_{2,3}^{(1)} < 1,~~\gamma_{2,3}^{(2)} + \gamma_{4}^{(2)} < 1.
\end{align}
 It can be {shown} that eliminating $\{\gamma_{1}^{(1)}, \gamma_{2,3}^{(1)}, \gamma_{4}^{(2)},\gamma_{2,3}^{(2)}\}$ in the above bounds by Fourier-Motzkin elimination leads to the rate region ${\mathcal R}_{1}^{\text{eg\ref{example:DCC:fails}}}$. Moreover, it can be seen that the sum of composite rates $\gamma_{1}^{(1)}$ and $\gamma_{2,3}^{(1)}$ is constrained by the link capacity of sender ${\mathcal S}_1$, while the sum of composite rates $\gamma_{2,3}^{(2)}$ and $\gamma_{4}^{(2)}$ is constrained by the link capacity of sender ${\mathcal S}_2$, {since} each sender independently uses point-to-point compression to compress its composite messages. {Therefore, to achieve $R_1 = R_4 =1-\epsilon$, where $\epsilon >0$ is arbitrarily small, both $\gamma_{1}^{(1)}$ and $\gamma_{4}^{(2)}$ are set to $1-\epsilon_1$ for some $\epsilon_1 \in (0, \epsilon)$, which in turn forces $\gamma_{2,3}^{(1)} < \epsilon_1$ and $\gamma_{2,3}^{(2)} < \epsilon_1$ and leads to $R_2 < 2\epsilon_1$ and $ R_3 < 2\epsilon_1$.}

\par However, composite message ${\mathcal M}_{2,3}$ is available at both senders. {Instead of compressing ${\mathcal M}_{2,3}$ to different composite indices (i.e., $W_{2,3}^{(1)}$ and $W_{2,3}^{(2)}$)}, both senders can use the same composite index $W_{2,3}$. In this way, senders ${\mathcal S}_1$ and ${\mathcal S}_2$ share a common composite index $W_{2,3}$ at rate $\gamma_{2,3}$ bcu, and each has a private index, namely, $W_1$ at rate $\gamma_{1}$ bcu and $W_4$ at rate $\gamma_{4}$ bcu, respectively. {The two senders can then collaboratively send these composite indices to the receivers using Slepian-Wolf-Cover random binning as in CCC for Proposition~\ref{proposition:MCCC}. {Specialized from~\eqref{equ:composite:rate:conditions:a}, the following constraints on composite rates $\{\gamma_1,\gamma_{2,3}, \gamma_{4}\}$ must be met to ensure that all receivers can successfully decode all composite indices with vanishing probability of error}:}
\begin{align}
&\gamma_1 < 1,~\gamma_4 < 1, \\
&\gamma_1 + \gamma_{2,3} < 2,~\gamma_4 + \gamma_{2,3} < 2,~\gamma_1 + \gamma_{4} < 2.
\end{align}
Given the composite indices decoded, each receiver then decodes its requested message by nonunique simultaneous decoding, with decoding choice ${\mathcal D}_{1} =\{1\}$ and ${\mathcal D}_{j} =[1:4]\backslash {\mathcal A}_j$~for~$j=2,3,4$ as before. Following from~\eqref{equ:MCCC:rate:constraints2}-\eqref{equ:MCCC:rate:constraints}, the resultant rate region is characterized by
\begin{align}
&R_1 < \gamma_{1}, ~~R_2 < \gamma_{2,3}, \\
&R_3 < \gamma_{2,3},~~R_4 < \gamma_{4},\\
&\gamma_1 < 1,~\gamma_4 < 1, \\
&\gamma_1 + \gamma_{2,3} < 2,~\gamma_4 + \gamma_{2,3} < 2,~\gamma_1 + \gamma_{4} < 2.
\end{align}
{It can be seen that for any arbitrary small $\epsilon >0$, the symmetric-rate tuple $(R_j =1-\epsilon, j\in [1:4])$ is now achievable, since all composite rates $\{\gamma_{1}, \gamma_{2,3}, \gamma_{4}\}$ can be set to $1-\epsilon_1$ simultaneously for some $\epsilon_1 \in (0, \epsilon)$}. Moreover, it can be further shown that eliminating $\{\gamma_{1}, \gamma_{2,3}, \gamma_{4}\}$ in the above bounds by Fourier-Motzkin elimination (all variables involved are non-negative) leads to a new achievable rate region, which agrees with the outer bound ${\mathcal R}_{\text{out}}^{\text{eg\ref{example:DCC:fails}}}$ of~\eqref{equ:eg2:outerbound}, thus establishing the capacity region. This example hence confirms the benefit of cooperative compression in CCC.

\begin{remark}
We have shown~\cite{li2017improved} that rate tuple $(R_j =1, j\in [1:4])$ can also be achieved by a linear index code for Example~\ref{example:DCC:fails}. Specifically, ${\mathcal S}_1$ sends $c_1= M_1 \oplus M_2 \oplus M_3$, while ${\mathcal S}_2$ sends $c_2 = M_2 \oplus M_3 \oplus M_4$. In this way, using side information, receiver~$2$ and receiver~$3$  each can decode their respective requested messages from $c_1$, while receiver~$4$ can decode $M_4$ from $c_2$. For receiver~$1$, knowing $M_4$, it can recover an XOR-message $w_{2,3} = M_2 \oplus M_3$ from $c_2$ and then cancel out $w_{2,3}$ in $c_1$ to decode $M_1$. Having the same linear function of $M_2$ and $M_3$ at both senders is crucial to achieve this rate tuple. This observation has motivated the proposed cooperative compression in CCC, where a common composite message at senders is encoded to the same composite index. For Example~\ref{example:DCC:fails}, the common composite index is $W_{2,3}^{(1)} = W_{2,3}^{(2)}$.~\remarkend
\end{remark}

\par We also revisit a 15-sender 4-message index-coding instance in Sadeghi~\textit{et~al.}~\cite{sadeghi2016distributed}.
\begin{example}[{an instance where ${\mathcal R} \subsetneq \{{\mathcal R}_{\textup{S}}, {\mathcal R}_{\textup{C}}\}$ and $\overline{{\mathcal R}_{\textup{S}}}=\overline{{\mathcal R}_{\textup{C}}} = {\mathcal C}$}]\label{example:parastoo}
In this example, there are
\begin{itemize}
  \item $K=15$ senders, with indices of messages ${\mathcal S}_1=\{1\}$, ${\mathcal S}_2=\{2\}$, ${\mathcal S}_3=\{3\}$, ${\mathcal S}_4=\{4\}$, ${\mathcal S}_5=\{1,2\}$, ${\mathcal S}_6=\{1,3\}$, ${\mathcal S}_7=\{3,4\}$, ${\mathcal S}_8=\{2,3\}$, ${\mathcal S}_9=\{2,4\}$, ${\mathcal S}_{10}=\{1,4\}$, ${\mathcal S}_{11}=\{1,2,3\}$, ${\mathcal S}_{12}=\{1,2,4\}$, ${\mathcal S}_{13}=\{1,3,4\}$, ${\mathcal S}_{14}=\{2,3,4\}$, ${\mathcal S}_{15}=\{1,2,3,4\}$, and each with link capacity $C_k =1$~bcu, $k \in [1:15]$.
  \item $N=4$ receivers, with side-information ${\mathcal A}_1 = \{4\}$, ${\mathcal A}_2 = \{3,4\}$, ${\mathcal A}_3 = \{1,2\}$, and ${\mathcal A}_4 = \{2,3\}$, respectively.
\end{itemize}
\end{example}
For this example, with decoding choice ${\mathcal D}_{1} =\{1\}$ and ${\mathcal D}_{j} =[1:4]\backslash {\mathcal A}_j$~for~$j=2,3,4$, {we establish that CCC with cooperative compression and without sender partitioning} leads to the following rate region
\begin{align}
 {\mathcal R}_{\text{C}}^{\text{eg\ref{example:parastoo}}}  = \left\{ \begin{array}{l}
 \left( {R_1 ,R_2 ,R_3 ,R_4 } \right) \in {\mathbb R}^4_+: \\
R_1 < 8,~R_2 < 8,~R_3 < 8,~R_4 < 8,\\
R_1 + R_2 < 12,~R_1 + R_3 < 12, \\
R_1 + R_4 < 12,~R_3 + R_4 < 12, \\
R_1 + R_2 + R_3 < 18.
 \end{array} \right\},
\end{align}
which strictly enlarges the previously reported region by partitioned DCC~\cite{sadeghi2016distributed}. Note that the same region here can also be attained by the use of a better sender-grouping (e.g., the second and fifth groups in~\cite[Table II]{sadeghi2016distributed} are combined as a single group) in partitioned DCC. We further prove that \textit{{the closure} of the new achievable rate region (i.e., $\overline{{\mathcal R}_{\textup{C}}^{\textup{eg\ref{example:parastoo}}}}$) is in fact the capacity region}. The converse proof requires a set of customized Shannon-type inequalities, since the existing outer bounds (the MAIS bound of~\eqref{equ:MAIS:outerbound}, and the polymatroidal bound~\cite{sadeghi2016distributed}) are both loose for this example. Details on the converse proof are deferred to Appendix~\ref{appendix:converse:parastoo}.

\begin{figure*}[t]
\centering
\includegraphics[width=0.9\textwidth]{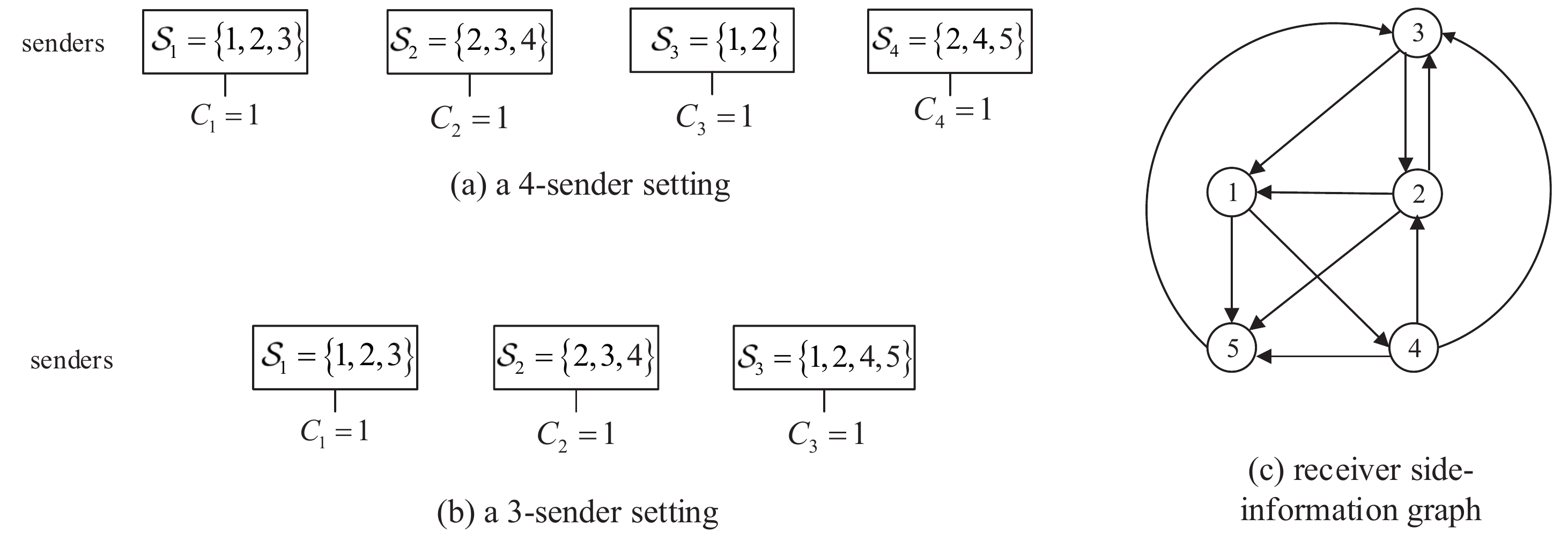}
\caption{Sender and receiver settings for the examples with 5-message: (a) a 4-sender setting in \textit{Example~\ref{example:MCCC1}}; (b) a 3-sender setting in \textit{Example~\ref{example:MCCC2}}; (c) receiver side-information graph considered in both \textit{Example~\ref{example:MCCC1}} and \textit{Example~\ref{example:MCCC2}}.} \label{fig:5Node:example}
\end{figure*}

\subsection{The Benefit of Sender Partitioning and Joint Link-and-Sender Partitioning}\label{sub:sec:benefit:link}
\par We first demonstrate the benefit of sender partitioning in CCC. We consider a 4-sender 5-message index-coding example as defined below.
\begin{example}[{an instance where ${\mathcal R}_{\textup{C}} \subsetneq {{\mathcal R}_{\textup{CS}}}$ and $ \overline{{\mathcal R}_{\textup{CS}}}= {\mathcal C}$}]\label{example:MCCC1}
In this example, there are
\begin{itemize}
  \item $K=4$ senders, with indices of messages ${\mathcal S}_1=\{1,2,3\}$, ${\mathcal S}_2=\{2,3,4\}$, ${\mathcal S}_3=\{1,2\}$, and ${\mathcal S}_4=\{2,4,5\}$, and each with link capacity $C_k =1$~bcu, $k\in[1:4]$;
  \item $N=5$ receivers, with side-information ${\mathcal A}_1 = \{4,5\}$, ${\mathcal A}_2 = \{1,3,5\}$, ${\mathcal A}_3 = \{1,2\}$, ${\mathcal A}_4 = \{2,3,5\}$ and ${\mathcal A}_5 = \{3\}$, respectively.
\end{itemize}
\end{example}
The above sender and receiver settings are depicted in Figs.~\ref{fig:5Node:example} (a) and (c), respectively.

\par In this instance, note that if we only consider the first two senders and the messages/receivers involved, the resulting index-coding subproblem coincides with the problem in \textit{Example~\ref{example:DCC:fails}}. As discussed earlier, cooperative compression is the key to achieve the capacity region for the problem. Moreover, in terms of optimal decoding choice, receiver~$1$ is restricted to decode only $M_1$. On the other hand, if we only consider the last two senders and the messages/receivers involved, we observe that {DCC-a (also CCC without sender partitioning)} suffices to attain the capacity region for the corresponding subproblem, but it requires receiver~$1$ to at least decode both $M_1$ and $M_2$. There is clearly a conflict of decoding requirements for receiver~$1$ between these two subproblems (sender groups) said. {These observations imply that} sender partitioning in CCC is crucial to achieve the capacity region for the whole problem.

\par {Specifically, by applying Proposition~\ref{proposition:MCCC} to this index-coding instance, if all senders are restricted to be in the same group, the best possible achievable rate region ${\mathcal R}_{\text{C}}^{\text{eg\ref{example:MCCC1}}}$} is achieved time-sharing between two different decoding choices: \textit{i}) ${\mathcal D}_{1} =\{1\}$, ${\mathcal D}_{j} =[1:5]\backslash {\mathcal A}_j, j\in[2:5]$; and \textit{ii}) ${\mathcal D}_{j} =[1:5]\backslash {\mathcal A}_j, j\in[1:4]$, ${\mathcal D}_{5} =\{5\}$. However, it is still not capacity-achieving. Instead, {CCC with senders $\{{\mathcal S}_1,{\mathcal S}_2\}$ as a group and $\{{\mathcal S}_3,{\mathcal S}_4\}$ as another group (i.e., $\tilde \Pi = \left\{\{1,2\}, \{3,4\}\right\}$, see Remark~\ref{remark:CCC:special:cases})} leads to the following achievable rate region:
\begin{align}
 {\mathcal R}_{\text{CS}}^{\text{eg\ref{example:MCCC1}}}  = \left\{ \begin{array}{l}
 \left( {R_1 ,R_2 ,R_3 ,R_4, R_5} \right) \in {\mathbb R}^5_+: \\
R_1 < 2,~R_3 < 2,~R_5 < 1,\\
R_1 + R_3 < 3,~R_4 + R_5 < 2, \\
R_1 + R_2 + R_5 < 4, \\
R_2 + R_4 + R_5 < 4, \\
R_3 + R_4 + R_5 < 3.
 \end{array} \right\},
\end{align}
which agrees with the MAIS outer bound~\eqref{equ:MAIS:outerbound} specialized to this index-coding problem, thus establishing the capacity region. Note that this example can also be used to verify that ${\mathcal R}_{\text{S}} \subsetneq {\mathcal R}_{\text{CS}}$ as depicted in Fig.~\ref{fig:scheme:comparison}, due to the necessity of cooperative compression.

\par We now demonstrate the benefit of joint link-and-sender partitioning in CCC. In particular, we study an example with the same receiver side-information graph as that in \textit{Example~\ref{example:MCCC1}}, but with only 3 senders as defined below.
\begin{example}[{an instance where ${\mathcal R}_{\textup{CS}}\subsetneq {\mathcal R}_{\textup{CLS}}$ and $\overline{{\mathcal R}_{\textup{CLS}}} = {\mathcal C}$}]\label{example:MCCC2}
In this example, there are
\begin{itemize}
  \item $K=3$ senders, with indices of messages ${\mathcal S}_1=\{1,2,3\}$, ${\mathcal S}_2=\{2,3,4\}$, and ${\mathcal S}_3=\{1,2,4,5\}$, and each with link capacity $C_k =1$~bcu, $k\in[1:3]$;
  \item $N=5$ receivers, with side-information ${\mathcal A}_1 = \{4,5\}$, ${\mathcal A}_2 = \{1,3,5\}$, ${\mathcal A}_3 = \{1,2\}$, ${\mathcal A}_4 = \{2,3,5\}$ and ${\mathcal A}_5 = \{3\}$, respectively.
\end{itemize}
\end{example}
The sender and receiver setting are illustrated in Figs.~\ref{fig:5Node:example} (b) and (c), respectively.
\par For this example, the MAIS outer bound of~\eqref{equ:MAIS:outerbound} is specialized to
\begin{align} \label{equ:MCCC2:outerbound}
{\mathcal R}_{\text{out}}^{\text{eg\ref{example:MCCC2}}}  = \left\{ \begin{array}{l}
 \left( {R_1 ,R_2 ,R_3 ,R_4, R_5} \right) \in {\mathbb R}^5_+  : \\
 R_3  \le 2,~~~~R_5  \le 1, \\
 R_1 + R_3  \le 3,~R_1  + R_5  \le 2,~R_4 + R_5 \le 2, \\
 R_1 + R_2 + R_5 \le 3,~R_1  + R_4 + R_5  \le 3, \\
 R_2 + R_4 + R_5 \le 3,~R_3 + R_4 + R_5 \le 3.
 \end{array} \right\}.
\end{align}
\par {For the inner bound, by Proposition~\ref{proposition:MCCC}}, {with all senders in a group, ${\mathcal R}_{\text{C}}^{\text{eg\ref{example:MCCC2}}}$ is achieved by time-sharing between two different decoding choices: \textit{i}) ${\mathcal D}_{1} =\{1\}$, ${\mathcal D}_{5} =\{5\}$, ${\mathcal D}_{j} =[1:5]\backslash {\mathcal A}_j, j\in[2:4]$; and \textit{ii}) ${\mathcal D}_{j} =[1:5]\backslash {\mathcal A}_j, j\in[1:4]$, ${\mathcal D}_{5} =\{5\}$, and it is characterized by:
\begin{align}
 {\mathcal R}_{\text{C}}^{\text{eg\ref{example:MCCC2}}}  = \left\{ \begin{array}{l}
 \left( {R_1 ,R_2 ,R_3 ,R_4, R_5} \right) \in {\mathbb R}^5_+: \\
 R_3  < 2,~~~~R_5  < 1, \\
 R_1 + R_3  < 3,~R_1  + R_5  < 2,~R_4 + R_5 < 2, \\
 R_1 + R_2 + R_5 < 3,~R_1  + R_4 + R_5  < 3, \\
 R_2 + R_4 + R_5 < 3,~R_3 + R_4 + R_5 < 3,\\
 2R_1 + R_2 + R_3 + R_4 + 2R_5 < 7,\\
 4R_1 + 3R_2 + R_3 + 3R_4 + 4R_5 < 15,\\
 3R_1 + 3R_2 + R_3 + 4R_4 + 4R_5 <  15.
 \end{array} \right\}.
\end{align}
In addition, by considering all possible non-trivial sender partitions of three senders, CCC with the best sender-partition $\tilde \Pi^* = \{{\tilde P}_1 = \{1,2\},{\tilde P}_2 = \{3\}\}$ leads to the following rate region:
\begin{align}
 &{\mathcal R}_{\text{CS}}^{\text{eg\ref{example:MCCC2}}}(\tilde \Pi^*)  =\nonumber \\
 &\left\{ \begin{array}{l}
 \left( {R_1 ,R_2 ,R_3 ,R_4, R_5} \right) \in {\mathbb R}^5_+: \\
 R_3  < 2,~~~~R_5  < 1, \\
 R_1 + R_3  < 3,~R_1  + R_5 < 2,~R_4 + R_5 < 2, \\
 R_1 + R_2 + R_5 < 3,~R_1  + R_4 + R_5  < 3, \\
 R_2 + R_4 + R_5 < 3,~R_3 + R_4 + R_5 < 3,\\
 R_1 + 2R_3 + R_4 + R_5 < 5.
 \end{array} \right\},
\end{align}
where relevant decoding sets are: \textit{i}) ${\mathcal D}_{1,{\tilde P}_1} =\{1\}, {\mathcal D}_{j,{\tilde P}_1} ={\mathcal S}_{{\tilde P}_1}\backslash {\mathcal A}_{j,{\tilde P}_1}$ for $j=2,3,4$; \textit{ii}) ${\mathcal D}_{j,{\tilde P}_2} ={\mathcal S}_{{\tilde P}_2}\backslash {\mathcal A}_{j,{\tilde P}_2}$ for $j=1,2,4$, and ${\mathcal D}_{5,{\tilde P}_2} =\{5\}$.
\par {Since none of the rate regions ${\mathcal R}_{\text{C}}^{\text{eg\ref{example:MCCC2}}}$ and ${\mathcal R}_{\text{CS}}^{\text{eg\ref{example:MCCC2}}}(\tilde \Pi^*)$ strictly contains the other, we thus have ${\mathcal R}_{\text{CS}}^{\text{eg\ref{example:MCCC2}}} = \text{conv}({\mathcal R}_{\text{CS}}^{\text{eg\ref{example:MCCC2}}}(\tilde \Pi^*) \bigcup {\mathcal R}_{\text{C}}^{\text{eg\ref{example:MCCC2}}})$. It can be checked that there is still a gap between the inner bound ${\mathcal R}_{\text{CS}}^{\text{eg\ref{example:MCCC2}}}$ and the outer bound ${\mathcal R}_{\text{out}}^{\text{eg\ref{example:MCCC2}}}$ of~\eqref{equ:MCCC2:outerbound}. For instance, the rate tuple ${\mathbf R}_c = (R_1 = 1.5, R_2 = 1, R_3 = 1.5, R_4 = 1, R_5 = 0.5)$ is in ${\mathcal R}_{\text{out}}^{\text{eg\ref{example:MCCC2}}}$, but it is not in the closure of ${\mathcal R}_{\text{CS}}^{\text{eg\ref{example:MCCC2}}}$. To close the gap, the most general CCC scheme with joint link-and-sender partitioning is required.}

\begin{table*}[t]
  \centering
  \caption{The rate regions attained by individual groups and the final combined rate region for Example~\ref{example:MCCC2}}
    \begin{tabular}{|c|c|c|}
    \hline
    {Link-Sender Groups} & {Achievable Rate regions} & {Optimal Decoding Sets} \bigstrut\\
    \hline
    {$({\tilde P}_1  = \{1,2\}, {\mathbf C}_{{\tilde P}_1} =[C_{1,{\tilde P}_1}, C_{2,{\tilde P}_1}])$} & {$
\begin{array}{c}
 ~\\
 R_{1,{\tilde P}_1 }  < C_{1,{\tilde P}_1},~~R_{4,{\tilde P}_1} < C_{2,{\tilde P}_1},\\
 R_{1,{\tilde P}_1} + R_{2,{\tilde P}_1 } < C_{1,{\tilde P}_1} + C_{2,{\tilde P}_1},\\
 R_{1,{\tilde P}_1} + R_{3,{\tilde P}_1 } < C_{1,{\tilde P}_1} + C_{2,{\tilde P}_1},\\
 R_{2,{\tilde P}_1} + R_{4,{\tilde P}_1 } < C_{1,{\tilde P}_1} + C_{2,{\tilde P}_1},\\
 R_{3,{\tilde P}_1} + R_{4,{\tilde P}_1 } < C_{1,{\tilde P}_1} + C_{2,{\tilde P}_1}.\\
 ~\\
 \end{array}$} &{$\begin{array}{c}
 {\mathcal D}_{1,{\tilde P}_1}=\{1\};\\
 {\mathcal D}_{j,{\tilde P}_1} ={\mathcal S}_{{\tilde P}_1}\backslash {\mathcal A}_{j,{\tilde P}_1},\\
 j=2,3,4.
 \end{array}$} \bigstrut\\
    \hline
    $({\tilde P}_2  = \{2,3\}, {\mathbf C}_{{\tilde P}_2} =[C_{2,{\tilde P}_2}, C_{3,{\tilde P}_2}])$ & {$\begin{array}{c}
 ~\\
 R_{3,{\tilde P}_2} < C_{2,{\tilde P}_2},~R_{1,{\tilde P}_2} + R_{5,{\tilde P}_2} < C_{3,{\tilde P}_2}, \\
 R_{1,{\tilde P}_2} + R_{2,{\tilde P}_2} + R_{5,{\tilde P}_2}  < C_{2,{\tilde P}_2} + C_{3,{\tilde P}_2}, \\
 R_{1,{\tilde P}_2} + R_{4,{\tilde P}_2} + R_{5,{\tilde P}_2}  < C_{2,{\tilde P}_2} + C_{3,{\tilde P}_2}, \\
 R_{2,{\tilde P}_2} + R_{4,{\tilde P}_2} + R_{5,{\tilde P}_2}  < C_{2,{\tilde P}_2} + C_{3,{\tilde P}_2}, \\
 R_{3,{\tilde P}_2} + R_{4,{\tilde P}_2} + R_{5,{\tilde P}_2}  < C_{2,{\tilde P}_2} + C_{3,{\tilde P}_2}.\\
 ~\\
 \end{array}$} & {$\begin{array}{c}
 {\mathcal D}_{1,{\tilde P}_2}=\{1\}, {\mathcal D}_{5,{\tilde P}_2}=\{5\};\\
 {\mathcal D}_{j,{\tilde P}_2} ={\mathcal S}_{{\tilde P}_2}\backslash {\mathcal A}_{j,{\tilde P}_2},\\
 j=2,3,4.
 \end{array}$} \bigstrut\\
 \hline
     {$({\tilde P}_3  = \{1,3\},{\mathbf C}_{{\tilde P}_3} =[C_{1,{\tilde P}_3}, C_{3,{\tilde P}_3}])$} & {$
\begin{array}{c}
 ~\\
 R_{3,{\tilde P}_3}  < C_{1,{\tilde P}_3},\\
 R_{1,{\tilde P}_3} + R_{3,{\tilde P}_3} < C_{1,{\tilde P}_3} + C_{3,{\tilde P}_3},\\
 R_{4,{\tilde P}_3} + R_{5,{\tilde P}_3} < C_{3,{\tilde P}_3},\\
 R_{1,{\tilde P}_3 } + R_{2,{\tilde P}_3} + R_{5,{\tilde P}_3} < C_{1,{\tilde P}_3} + C_{3,{\tilde P}_3},\\
 R_{1,{\tilde P}_3 } + R_{4,{\tilde P}_3} + R_{5,{\tilde P}_3} < C_{1,{\tilde P}_3} + C_{3,{\tilde P}_3},\\
 R_{2,{\tilde P}_3 } + R_{4,{\tilde P}_3} + R_{5,{\tilde P}_3} < C_{1,{\tilde P}_3} + C_{3,{\tilde P}_3}.\\
 ~\\
 \end{array}$} &$\begin{array}{c}
 {\mathcal D}_{1,{\tilde P}_3}=\{1\}, {\mathcal D}_{5,{\tilde P}_3}=\{5\};\\
 {\mathcal D}_{j,{\tilde P}_3} ={\mathcal S}_{{\tilde P}_3}\backslash {\mathcal A}_{j,{\tilde P}_3},\\
 j=2,3,4.
 \end{array}$ \bigstrut\\
    \hline
    {$\begin{array}{c}
    ~\\
    \text{Combined~Region}~{\mathcal R}_{\text{CLS}}^{\text{eg\ref{example:MCCC2}}}(\tilde \Pi,{\mathbf C}_{\tilde \Pi})\\
    (R_j= \sum\nolimits_{k=1}^{3} R_{j,{\tilde P}_k}, j \in [1:4],\\
    R_5 = R_{5,{\tilde P}_2} + R_{5,{\tilde P}_3},~~\text{s.t.},\\
    C_{1,{\tilde P}_1} + C_{1,{\tilde P}_3} \le 1,\\
    C_{2,{\tilde P}_1} + C_{2,{\tilde P}_2} \le 1,\\
    C_{3,{\tilde P}_2} + C_{3,{\tilde P}_3} \le 1.)\\
    ~\\
    \end{array}$ } & \multicolumn{2}{c|}{$\begin{array}{c}
 R_3  < 2,~~~~R_5  < 1, \\
 R_1 + R_3  < 3,~R_1  + R_5  < 2,~R_4 + R_5 < 2, \\
 R_1 + R_2 + R_5 < 3,~R_1  + R_4 + R_5  < 3, \\
 R_2 + R_4 + R_5 < 3,~R_3 + R_4 + R_5 < 3.
 \end{array}$} \bigstrut\\
    \hline
    \end{tabular}%
  \label{tab:rate:eg5}%
\end{table*}%

\par {To show this, in the evaluation of inner bound attained by CCC, consider $\tilde \Pi = \{{\tilde P}_1=\{1,2\},{\tilde P}_2=\{2,3\},{\tilde P}_3=\{1,3\}\}$ with three pair-wise sender subsets. This implies that each sender will split its link capacity $C_k$ into two non-negative portions and allocate each portion to the transmission in a different sender subset. The following three link-sender groups are formed for the index-coding example studied: $({\tilde P}_1  = \{1,2\}, {\mathbf C}_{{\tilde P}_1} =[C_{1,{\tilde P}_1}, C_{2,{\tilde P}_1}])$, $({\tilde P}_2  = \{2,3\}, {\mathbf C}_{{\tilde P}_2} =[C_{2,{\tilde P}_2}, C_{3,{\tilde P}_2}])$ and $({\tilde P}_3  = \{1,3\},{\mathbf C}_{{\tilde P}_3} =[C_{1,{\tilde P}_3}, C_{3,{\tilde P}_3}])$, s.t., $C_{1,{\tilde P}_1} + C_{1,{\tilde P}_3} \le 1$, $C_{2,{\tilde P}_1} + C_{2,{\tilde P}_2} \le 1$ and $C_{3,{\tilde P}_2} + C_{3,{\tilde P}_3} \le 1$.} By this link-sender group construction and invoking Proposition~\ref{proposition:MCCC}, we thus can first obtain the individual rate region under each link-sender group trough~\eqref{equ:MCCC}--\eqref{equ:MCCC:rate}, and then obtain the final combined rate region ${\mathcal R}_{\text{CLS}}^{\text{eg\ref{example:MCCC2}}}(\tilde \Pi, {\mathbf C}_{\tilde \Pi})$ via~\eqref{equ:MCCC:FM1}. Details on the optimal decoding sets and the individual rate regions are provided in Table~\ref{tab:rate:eg5}. We note that ${\mathcal R}_{\text{CLS}}^{\text{eg\ref{example:MCCC2}}}(\tilde \Pi, {\mathbf C}_{\tilde \Pi})$ agrees with the MAIS outer bound ${\mathcal R}_{\text{out}}^{\text{eg\ref{example:MCCC2}}}$ in form of~\eqref{equ:MCCC2:outerbound}, thus establishing the capacity region. {This example demonstrates the benefit of joint link-and-sender partitioning in CCC, and it confirms that ${\mathcal R}_{\text{CS}} \subsetneq {\mathcal R}_{\text{CLS}}$ as in Fig.~\ref{fig:scheme:comparison}.}

\section{{Comparisons and Discussions}}\label{sec:discussions}

\par We now discuss and compare CCC with two enhanced distributed composite-coding schemes without cooperative composite-message compression presented in recent works.

\par The first enhanced scheme replaces sender-partitioning in partitioned DCC with joint link-and-sender partitioning as in CCC, and it is called a modified DCC (mDCC) scheme~\cite{li2017improved} (due to the same authors of this paper). The achievable rate region under arbitrary admissible joint link-and-sender partition is denoted by ${\mathcal R}_{\text{LS}}(\tilde \Pi)$ and is characterized in~\cite[Proposition 3]{li2017improved}. The general achievable rate region ${\mathcal R}_{\text{LS}}$ is given by the {convex hull} of all ${\mathcal R}_{\text{LS}}(\tilde \Pi)$'s under all possible $\tilde \Pi$'s, similar to~\eqref{equ:convexhull:RCLS}. In general, ${\mathcal R}_{\text S} \subseteq {\mathcal R}_{\text{LS}}$, and the inclusion can be strict~\cite[Example 1]{li2017improved}. However, mDCC is still suboptimal as compared with CCC (i.e., ${\mathcal R}_{\text{LS}} \subsetneq {\mathcal R}_{\text{CLS}}$), which can be shown via \textit{Example~\ref{example:MCCC2}} due to the necessity of cooperative compression in achieving the full capacity region.

\par Liu~\textit{et~al.} proposed another enhanced scheme, called fractional DCC, in an independent and parallel work~\cite[Section IV]{liu2017distributed}. Fractional DCC includes the concept of joint link-and-sender partitioning we proposed and also includes a new strategy to accommodate more flexible composite-rate allocation among different decoding choices than time sharing. Hence, fractional DCC improves upon mDCC in general. However, this scheme, without cooperative compression of composite messages, is strictly suboptimal for the two-sender index-coding instance in \textit{Example~\ref{example:DCC:fails}}, while CCC achieves the capacity region for this instance. Specifically, the best possible rate region attained by fractional DCC (two senders as a group) for \textit{Example~\ref{example:DCC:fails}} is
\begin{align}
 {\mathcal R}_{\text{Fractional DCC}}^{\text{eg\ref{example:DCC:fails}}}  = \left\{ \begin{array}{l}
 \left( {R_1 ,R_2 ,R_3 ,R_4 } \right) \in {\mathbb R}^4_+: \\
 R_1  < 1,~~~~R_4  < 1, \\
R_1  + R_2  < 2,~R_1  + R_3  < 2,\\
R_2  + R_4  < 2,~R_3  + R_4  < 2,\\
2R_1 + R_2 + R_3 + R_4 < 4.\\
\end{array} \right\},
\end{align}
which does not coincide with the MAIS outer bound of~\eqref{equ:eg2:outerbound}. The exact performance relationship between fractional DCC and CCC remains open in general.

\section{Conclusions}\label{sec:conclusions}
\par In this paper, we have developed a new achievable scheme via a random coding approach and established new bounds on the capacity region for the multi-sender unicast index-coding problem. In particular, we have introduced a cooperative compression of composite messages in multi-sender composite coding to leverage potential overlapping of messages at different senders to support larger composite rates than partitioned DCC~\cite{sadeghi2016distributed}. We have also introduced a joint link-and-sender partitioning that generalizes existing sender partitioning. By the combined use of the new techniques proposed, we have devised a new multi-sender CCC scheme, which improves upon partitioned DCC in general. It can seen that compared with single-sender index coding, the multi-sender index-coding problem is richer and more difficult to solve, due to the varying availability of messages and the nature of distributed encoding at each sender. The current work hence serves as an intermediate step towards a full understanding of the problem.
\par For future work, we will investigate whether or not CCC proposed here suffices to achieve the capacity region for all non-isomorphic 4- or 5-message index-coding instances with arbitrary admissible sender setting and with arbitrary link capacities, in the same spirit of the study to all 3-message index-coding instances by Sadeghi~\textit{et~al.}~\cite{sadeghi2016distributed}. Our preliminary studies focused on the 4-message case and considered all 15 senders activated each with unit link capacity as in~\cite{sadeghi2016distributed,liu2017distributed}. It was found that: (\textit{i}) there are $103$ out of $218$ non-isomorphic side-information digraphs (see~\cite{liu2017distributed} for a list of these $218$ digraphs), for which the CCC-induced inner bound ${\mathcal R}_{\text{C}}$ agrees with the MAIS outer bound, thus establishing the capacity regions for these instances; (\textit{ii}) there are $7$ side-information digraphs (including the one in \textit{Example~\ref{example:parastoo}}), for which the CCC-induced inner bound ${\mathcal R}_{\text{C}}$ agrees with a customized outer bound similar to that derived for \textit{Example~\ref{example:parastoo}}, thus establishing their capacity regions. The indices of these solved instances are listed in Appendix~\ref{appendix:problems:solved}. For the remaining $108$ digraphs, no firm conclusion was reached yet. To completely solve them, we might need customized link-sender partitioning optimization for each instance in CCC, new advanced coding schemes, or new customized outer bounds, in which case it is highly non-trivial and requires significant extra efforts.
\par {Another line of future work is to investigate the optimal broadcast rate for a general multi-sender unicast index-coding problem. Thapa \textit{et al.}~\cite{thapa2016twosender} has studied a two-sender setup as an intermediate step and generalized existing cycle-cover, clique-cover and local-chromatic approaches to solve the two-sender problem. More graph-based approaches, such as (local) partial clique cover~\cite{birk1998informed,agarwal2016local} and min-rank~\cite{bar2011index,shanmugam2014graph} are also worth considering. Random coding approach such as CCC proposed here can also be applied to establish upper bounds on the broadcast rate.}

\appendices
\section{An Alternative Proof of the Multi-Sender MAIS Outer Bound} \label{appendix:MAIS}
Recall from the multi-sender index code~\textit{Definition~\ref{def:index:code}}, each message $M_j$ is independently and uniformly distributed over the set $[1:2^{nR_j}]$, where $n$ is the number of channel uses, and $L_k$ denotes the output of each sender ${\mathcal S}_k$. Let ${\mathbf L} = \{L_1,L_2,\cdots, L_{K}\}$ denote the collection of outputs at all senders, let ${\mathbf L}_{{\mathcal M}_S} \subseteq {\mathbf L}$ denote the collection of outputs that are completely determined by messages only in ${{\mathcal M}_S}$, and let $S^c$ denote the complement of $S$ with respect to $\left[1:N\right]$. For any $S \subseteq \left[1:N\right]$ such that the subgraph of $G$ induced by $S$ is acyclic, we have:
\begin{align}
n\sum\nolimits_{j \in S} {R_j } &= H({\mathcal M}_{S}) \label{equ:MAIS:1}\\
&= H({\mathcal M}_{S} \left| {\mathcal M}_{S^c}\right.) \label{equ:MAIS:2}\\
&= H({\mathbf L}, {\mathcal M}_{S} \left| {\mathcal M}_{S^c}\right.) \label{equ:MAIS:3}\\
&= H({\mathbf L}\left| {\mathcal M}_{S^c}\right.) \label{equ:MAIS:5}\\
&\le H({\mathbf L}\backslash {\mathbf L}_{{\mathcal M}_{S^c}} ) \label{equ:MAIS:6} \\
&\le n\sum\limits_{k \in [1:K] :~{\mathcal S}_k  \cap S \ne \emptyset} {C_k}, \label{equ:MAIS:7}
\end{align}
where equality~\eqref{equ:MAIS:2} holds due to the independence of messages; equality~\eqref{equ:MAIS:3} follows from the encoding functions of~\eqref{equ:def:encodings}, while equality~\eqref{equ:MAIS:5} holds because $H({\mathcal M}_{S}\left|{\mathbf L},{\mathcal M}_{S^c}\right.)=0$, which is deduced by the facts that the subgraph of $G$ induced by $S$ is acyclic and thus each message $M_j, j\in S$ can be always decoded in a certain order according to the decoding functions of~\eqref{equ:def:decodings}, given ${\mathbf L},{\mathcal M}_{S^c}$ and the subset of messages recovered before $M_j$; \eqref{equ:MAIS:6}~holds because $H({\mathbf L}\left| {\mathcal M}_{S^c}\right.) = H({{\mathbf L}\backslash {\mathbf L}_{{\mathcal M}_{S^c}} }\left| {\mathcal M}_{S^c}\right.)$ and conditioning reduces entropy; \eqref{equ:MAIS:7}~holds because the output of sender ${\mathcal S}_k$ is at most $C_k$ bcu constrained by its link capacity.

\section{Proof of Proposition~\ref{proposition:CCC:over:DCC}}\label{appendix:proposition:CCC:over:DCC}
\par {We first prove ${\mathcal R}_{\text{S}} \subseteq {\mathcal R}_{\text{CS}}$. Consider any arbitrary sender partition $\Pi \in {\mathbf \Pi}_{\text{S}}$ in Proposition~\ref{proposition:DCC}, and assume the same sender partition is used in Proposition~\ref{proposition:MCCC}. In this case, we have ${\tilde P} = P \in \Pi$ and $C_{k,{\tilde P}} = C_{k}$, $\forall k \in {\tilde P}, \forall {\tilde P} \in \Pi$ in~\eqref{equ:MCCC:rate}.}

\par {Consider any sender-group ${P} \in \Pi$ and any subset $\tilde{\mathcal K} \subseteq {P}$. For convenience, define $f({\tilde {\mathcal K}}) = \{{\mathcal J}_2 \in {\mathcal I}_{{\tilde {\mathcal K}}}: {\mathcal J}_2 \notin {\mathcal I}_{{\tilde {\mathcal K}}^c},{\mathcal J}_2 \not\subseteq {\mathcal A}_{j,{P}}\}$ as a set of indices of composite rates that appear in~\eqref{equ:MCCC:rate}.(b). Similarly, define $h(k) =\{{\mathcal J}_2  \subseteq {\mathcal S}_k: {\mathcal J}_2  \not\subseteq {\mathcal A}_{j,{P}} \}$ as a set of indices of composite rates that appear in~\eqref{equ:DCC:mainproblem}.(b). In addition, given any fixed admissible ${\mathcal J} \in f(\tilde{\mathcal K})$, define $g({\mathcal J}) = \{k: k\in {P},{\mathcal J}\subseteq {\mathcal S}_k\}$, and let $\gamma_{{\mathcal J},{P}} = \sum\nolimits_{k\in g({\mathcal J})} \gamma_{{\mathcal J},{P}}^{(k)}$.}

\par {For the chosen $\tilde{\mathcal K}$ and ${\mathcal J}$, if $k \in g({\mathcal J})$, we must have $k \in \tilde{\mathcal K}$. We can prove this by contradiction. Specifically, if $k \in g({\mathcal J})$, then ${\mathcal J} \subseteq {\mathcal S}_k$ by the definition of $g({\mathcal J})$. This hence implies that ${\mathcal J} \in {\mathcal I}_{\{k\}}$ by the definition of ${\mathcal I}_{\{k\}}$. Now, suppose that $k \in \tilde{\mathcal K}^c$, then ${\mathcal J} \in {\mathcal I}_{\{k\}} \subseteq {\mathcal I}_{\tilde{\mathcal K}^c}$, which contradicts the assumption that ${\mathcal J} \in f(\tilde{\mathcal K})$.}

\par {We now inspect the constraints imposed on composite rates in~\eqref{equ:MCCC:rate} and~\eqref{equ:DCC:mainproblem}, respectively. For any fixed $\tilde{\mathcal K}$, constraint~\eqref{equ:MCCC:rate}.(b) reads as
\begin{align}
\sum\limits_{{\mathcal J} \in f(\tilde {\mathcal K})} \sum\limits_{k \in g({\mathcal J})}\gamma_{{\mathcal J},P}^{(k)}~<~\sum\limits_{k \in \tilde {\mathcal K}}C_{k}, \label{equ:mccc:constraint}
\end{align}
while constraint~\eqref{equ:DCC:mainproblem}.(b) implies that
\begin{align}
\sum\limits_{k\in \tilde {\mathcal K}}\sum\limits_{{\mathcal J'} \in h(k)} \gamma^{(k)}_{{\mathcal J'},{P}}~<~\sum\limits_{k \in \tilde {\mathcal K}}C_{k}.\label{equ:dcc:lp:constraint}
\end{align}
For each $({\mathcal J}, k)$ in the left-hand side of~\eqref{equ:mccc:constraint}, the corresponding $\gamma_{{\mathcal J},P}^{(k)}$ must appear in the left-hand side of~\eqref{equ:dcc:lp:constraint}, because: $\textit{i)}$ for any ${\mathcal J} \in f(\tilde{\mathcal K})$, if $k \in g({\mathcal J})$, then $k \in \tilde{\mathcal K}$ as proved before; and $\textit{ii)}$ if ${\mathcal J} \in f(\tilde{\mathcal K})$, then ${\mathcal J} \in h(k)$ for some $k\in \tilde{\mathcal K}$, as $f({\tilde {\mathcal K}}) \subseteq \bigcup\nolimits_{k \in {\tilde {\mathcal K}}}h(k)$. Since we only count distinct $({\mathcal J}, k)$'s, constraint~\eqref{equ:dcc:lp:constraint} must be more restrictive than constraint~\eqref{equ:mccc:constraint}. We therefore conclude that ${\mathcal R}_{\text{S}}(\Pi) \subseteq {\mathcal R}_{\text{CS}}(\Pi)$ in general. Since this is true for any $\Pi \in {\mathbf \Pi}_{\text{S}}$, we further have ${\mathcal R}_{\text{S}} \subseteq {\mathcal R}_{\text{CS}}$.}

\par {Noting ${\mathcal R}_{\text{CS}} \subseteq {\mathcal R}_{\text{CLS}}$ from~\eqref{equ:CCC:specicalcases}, we hence establish that ${\mathcal R}_{\text{S}} \subseteq {\mathcal R}_{\text{CLS}}$. The inclusion can be strict as shown by \textit{Example~\ref{example:MCCC1}} and \textit{Example~\ref{example:MCCC2}} in Section~\ref{sec:cooperative:compression}.}

\section{Converse Proof of the Capacity Region for Example~\ref{example:parastoo}}\label{appendix:converse:parastoo}
\par Recall from the multi-sender index code~\textit{Definition~\ref{def:index:code}}, each message $M_j$ is independently and uniformly distributed over the set $[1:2^{nR_j}]$, where $n$ is the number of channel uses. The output $L_k$ of each sender ${\mathcal S}_k$  is a function of the messages available at the sender. Thus, we have
\begin{align}
H(L_k \left|{\mathcal M}_{{\mathcal S}_k}\right.) = 0,~~k \in [1:15]. \label{equ:encodingfunctions}
\end{align}
Let ${\mathbf L} = \{L_1,L_2,\cdots, L_{15}\}$ denote the collection of all outputs at the $15$ senders. By the definition of decoders, we have the following decodability conditions at receivers:
\begin{align}
&H(M_1 \left|{\mathbf L}, M_4\right.) = 0, \label{equ:decodability:condition1}\\
&H(M_2 \left|{\mathbf L}, M_3, M_4\right.) = 0,\label{equ:decodability:condition2}\\
&H(M_3 \left|{\mathbf L}, M_1, M_2\right.) = 0, \label{equ:decodability:condition3}\\
&H(M_4 \left|{\mathbf L}, M_2, M_3\right.) = 0. \label{equ:decodability:condition4}
\end{align}
With these conditions, we now proceed with the converse proof.
\begin{enumerate}
    \item The outer bounds on each individual rate $R_j \le 8$ and on each rate pair concerned are included in the MAIS outer bound of~\eqref{equ:MAIS:outerbound}. The proof is hence omitted here.
    \item We now prove that $R_1 + R_2 + R_3 \le 18$. We first note that
      \begin{align}
       &n(R_1+R_2+R_3) \nonumber \\
       =&~H(M_1,M_2,M_3) \\
                      =&~H({\bf{L}},M_1,M_2,M_3{\left|M_4 \right.}) \label{equ:bound2:1}\\
                      =&~H({\bf{L}}{\left|M_4\right.}) + H(M_1,M_2,M_3{\left|M_4,{\bf{L}}\right.}) \label{equ:bound2:2}\\
                      \le&~H({\bf{L}}\backslash L_4) + H(M_1,M_2,M_3{\left|M_4,{\bf{L}}\right.}) \label{equ:bound2:3} \\
                      \le&~14n +  H(M_1,M_2{\left|M_4,{\bf{L}}\right.}) + H(M_3{\left|M_1,M_2,M_4,{\bf{L}}\right.}) \label{equ:bound2:4}\\
                      =&~14n + H(M_1,M_2{\left|M_4,{\bf{L}}\right.}) \label{equ:bound3:1} \\
                      =&~14n + H(M_1{\left|M_4,{\bf{L}}\right.}) + H(M_2{\left|L_1,L_4,{\bf{L}}\right.}) \\
                      =&~14n + H(M_2{\left|M_1,M_4,{\bf{L}}\right.}), \label{equ:bound3:2}
      \end{align}
      where \eqref{equ:bound2:1} follows the independence of messages and encoding functions of~\eqref{equ:encodingfunctions}, \eqref{equ:bound2:2} follows the chain rule of entropy, \eqref{equ:bound2:3} holds as $L_4$ is a function of $M_4$ and conditioning reduces entropy, \eqref{equ:bound3:1} holds due to the decodability condition~\eqref{equ:decodability:condition3} for $M_3$, and \eqref{equ:bound3:2} holds due to the decodability condition~\eqref{equ:decodability:condition1} for $M_1$.
      \par Now, suppose we can prove that
       \begin{align}
         H(M_2{\left|M_1,M_4,{\bf{L}}\right.}) \le 4n, \label{equ:target1}
       \end{align}
       then we are done.
      \par Towards this end, we examine the following two facts:
       \begin{itemize}
         \item $H(M_2{\left|M_1,M_4\right.}) $\\
           $\le H(L_2,L_{5},L_{8},L_{9},L_{11},L_{12},L_{14},L_{15}{\left|M_1,M_4\right.})$:
           \begin{align}
              &H(M_2{\left|M_1,M_4\right.}) \nonumber\\
              =&~H(M_2{\left|M_1,M_4,M_3\right.})\\
              =&~H(M_2,L_2,L_{5},L_{8},L_{9},L_{11},L_{12},L_{14},L_{15}| \nonumber \\
              &M_1,M_4,M_3,L_1,L_3,L_4,L_{6},L_{7},L_{10},L_{13}) \label{equ:bound4:1}\\
              =&~H(L_2,L_{5},L_{8},L_{9},L_{11},L_{12},L_{14},L_{15}| \nonumber\\
              &M_1,M_4,M_3,L_1,L_3,L_4,L_{6},L_{7},L_{10},L_{13})\nonumber \\
              &~~~~~~~~~~~~~~~+H(M_2{\left|M_1,M_4,M_3,{\bf L}\right.}) \label{equ:bound4:2}\\
              =&~H(L_2,L_{5},L_{8},L_{9},L_{11},L_{12},L_{14},L_{15}|\nonumber\\
              &M_1,M_4,M_3,L_1,L_3,L_4,L_{6},L_{7},L_{10},L_{13}) \label{equ:bound4:3}\\
              \le &~H(L_2,L_{5},L_{8},L_{9},L_{11},L_{12},L_{14},L_{15}{\left|M_1,M_4\right.}), \label{equ:bound4:4}
           \end{align}
           where \eqref{equ:bound4:1} follows the definition of encoding functions, \eqref{equ:bound4:2} uses the chain rule of entropy, \eqref{equ:bound4:3} holds as $H(M_2{\left|M_1,M_4,M_3,{\bf L}\right.})=0$ due to the decodability condition~\eqref{equ:decodability:condition2} for $M_2$, and \eqref{equ:bound4:4} holds because conditioning reduces entropy.
         \item $I(M_2;L_2,L_{5},L_{8},L_{9},L_{11},L_{12},L_{14},L_{15}{\left|M_1,M_4\right.})$ can be expanded in two ways:
             \begin{align}
              &I(M_2;L_2,L_{5},L_{8},L_{9},L_{11},L_{12},L_{14},L_{15}{\left|M_1,M_4\right.}) \nonumber\\
              =&~{H(M_2{\left|M_1,M_4\right.})}-H(M_2|M_1,M_4,L_2,L_{5},L_{8}, \nonumber \\
              &~~~~~~~~~~~~~~~~~~~~~~~~L_{9},L_{11},L_{12},L_{14},L_{15})\label{equ:bound5:2}\\
              =&~{H(L_2,L_{5},L_{8},L_{9},L_{11},L_{12},L_{14},L_{15}{\left|M_1,M_4\right.})}\nonumber\\
              &~-{H(L_2,L_{5},L_{8},L_{9},L_{11},L_{12},L_{14},L_{15}|} \nonumber\\
              &~~~~~~~~~~~~~~~~~~~~~~~~~~~~~~~~~~~M_1,M_2,M_4)\\
              =&~{H(L_2,L_{5},L_{8},L_{9},L_{11},L_{12},L_{14},L_{15}{\left|M_1,M_4\right.})}\nonumber\\
              &~-{H(L_{8},L_{11},L_{14},L_{15}{\left|M_1,M_2,M_4\right.})}, \label{equ:bound5:1}
             \end{align}
        where \eqref{equ:bound5:1} holds because ${L_2, L_{5}, L_{9},L_{12}}$ are deterministic functions of $\{M_1,M_2,M_4\}$.
        \par Now, using the first fact and by comparing~\eqref{equ:bound5:2} and \eqref{equ:bound5:1}, we can prove that the following inequality holds:
        \begin{align}
        &{H(M_2{\left|M_1,M_4,L_2,L_{5},L_{8},L_{9},L_{11},L_{12},L_{14},L_{15}\right.})}  \nonumber\\ &\le~{H(L_{8},L_{11},L_{14},L_{15}\left|M_1,M_2,M_4\right.}). \label{equ:key:ineq}
        \end{align}
       \end{itemize}
     \par Given this inequality, we thus have
      \begin{align}
      &H(M_2{\left|M_1,M_4,{\bf{L}}\right.}) \nonumber \\
        &\le {H(M_2{\left|M_1,M_4,L_2,L_{5},L_{8},L_{9},L_{11},L_{12},L_{14},L_{15}\right.})} \\
        &\le {H(L_{8},L_{11},L_{14},L_{15}{\left|M_1,M_2,M_4\right.})} \\
        &\le H(L_{8},L_{11},L_{14},L_{15}), \\
        &\le 4n. \label{equ:key:ineq2}
      \end{align}
  Finally, combining bounds \eqref{equ:bound3:2} and \eqref{equ:key:ineq2}, we conclude that
  \begin{align}
  R_1 + R_2 + R_3 \le 18,
  \end{align}
  which completes the proof for the outer bound on the capacity region.
\end{enumerate}

{\section{A List of Solved Multi-Sender Index-Coding Problem Instances with 4 Messages and 15 Senders Each with Unit Link Capacity}\label{appendix:problems:solved}
\begin{enumerate}[(i)]
  \item The indices of $103$ side-information digraphs, for which the CCC-induced inner bound ${\mathcal R}_{\text{C}}$ agrees with the MAIS outer bound, thus establishing the capacity region for each instance:
  \begin{align}
   &1, 2, 3, 5, 6, 7, 8, 10, 11, 12, 13, 15, 17, 19, 20, 22, 25, 26, \nonumber\\
   &33, 35, 38, 39, 40, 41, 42, 43, 44, 47, 49, 63, 65, 67, 69, 70,\nonumber \\
   &71, 72, 75, 76, 77, 78, 79, 83, 85, 100, 103, 105, 106, 107,\nonumber\\
   &108, 109, 110, 111, 117, 123, 125, 126, 127, 130, 131, 132,\nonumber\\
   &133, 142, 143, 144, 145, 152, 153, 154, 161, 163, 164, 165, \nonumber \\
   &166, 167, 168, 169, 174, 177, 183, 184, 185, 186, 187, 193, \nonumber \\
   &194, 195, 196, 197, 198, 201, 205, 206, 207, 208, 209, 210, \nonumber\\
   &211, 213, 214, 215, 216, 217, 218; \nonumber
  \end{align}
  \item The indices of $7$ side-information digraphs for which the CCC-induced inner bound ${\mathcal R}_{\text{C}}$ agrees with a customized outer bound similar to that derived for \textit{Example~\ref{example:parastoo}}, thus establishing the capacity region for each instance: $149, 155, 176, 179, 200, 203, 212$.
\end{enumerate}}

\section*{Acknowledgment}
The authors would like to thank the Associate Editor and the anonymous reviewers for their valuable comments that helped to improve the presentation
of this paper.

\bibliographystyle{IEEETran}

\end{document}